\documentclass[11pt]{article}

\usepackage[left=1in, right=1in, top=1in, bottom=1in]{geometry}
\usepackage[utf8]{inputenc}
\usepackage[T1]{fontenc}
\usepackage{bm}
\usepackage{type1cm}
\usepackage{lettrine}
\usepackage{amsmath,amssymb,amsthm}
\usepackage{moreverb}
\usepackage{mathtools}
\usepackage{amsmath}
\usepackage{svg}
\usepackage{amssymb}
\usepackage{algorithmic}
\usepackage{graphics}
\usepackage{graphicx}
\usepackage{subfigure}
\usepackage{caption}
\usepackage{extarrows}
\usepackage{color}
\usepackage{framed}
\usepackage{wrapfig}
\usepackage{algorithm}
\usepackage{algorithmic}
\usepackage{bm}
\usepackage{mathrsfs}
\usepackage{mathabx}
\usepackage{multirow}
\usepackage{longtable}
\usepackage{hyperref}
\usepackage{paralist}
\usepackage{indentfirst}
\usepackage{relsize}
\usepackage{extarrows}
\usepackage{upgreek}
\usepackage{bm}
\usepackage{mwe}
\usepackage{booktabs}
\usepackage{authblk}
\usepackage{lettrine}
\usepackage{type1cm}
\usepackage{threeparttable}
\usepackage[sort&compress,numbers]{natbib}
\usepackage[figurename=Figure]{caption}
\usepackage[normalem]{ulem}
\usepackage{adjustbox}
\usepackage{xcolor}

\usepackage{amsmath,amsfonts,bm}

\def\eqref#1{equation~\ref{#1}}

\def\1{\bm{1}}

\def\va{{\bm{a}}}

\def\vf{{\bm{f}}}

\def\vl{{\bm{l}}}

\def\vepsilon{{\bm{\epsilon}}}

\def\mA{{\bm{A}}}

\def\mF{{\bm{F}}}

\def\mL{{\bm{L}}}

\DeclareMathAlphabet{\mathsfit}{\encodingdefault}{\sfdefault}{m}{sl}
\SetMathAlphabet{\mathsfit}{bold}{\encodingdefault}{\sfdefault}{bx}{n}

\def\gM{{\mathcal{M}}}

\def\sR{{\mathbb{R}}}

\graphicspath{ {./Figure/} }
\usepackage[font=footnotesize,labelfont=bf]{caption}

\newcommand{\graph}[1]{\mathcal{}}
\providecommand{\keywords}[1]{\textbf{\textit{Keywords: }} #1}

\hypersetup{
bookmarks=true,
bookmarksopen=true,
bookmarksnumbered=true,
unicode=false,
pdftoolbar=true,
pdfmenubar=true,
pdffitwindow=false,
pdfstartview={FitH},
pdftitle={My title},
pdfauthor={Author},
pdfsubject={Subject},
pdfcreator={Creator},
pdfproducer={Producer},
pdfkeywords={keywords},
pdfnewwindow=true,
colorlinks=true,
linkcolor=blue,
citecolor=blue,
filecolor=blue,
urlcolor=blue
}

\begin{document}
\title{\textbf{InvDesFlow-AL: Active Learning-based Workflow for Inverse Design of Functional Materials }}

\author[1]{Xiao-Qi Han}
\author[1]{Peng-Jie Guo}
\author[1,*]{Ze-Feng Gao}
\author[2,*] {Hao Sun}
\author[1,3,*]{Zhong-Yi Lu}

\affil[1]{\small School of Physics, Renmin University of China, Beijing, China}
\affil[2]{\small Gaoling School of Artificial Intelligence, Renmin University of China, Beijing, China}
\affil[3]{\small School of Engineering Science, University of Chinese Academy of Sciences, Beijing, China\vspace{18pt}} 
\affil[*]{Corresponding authors\vspace{12pt}}

\date{}

\maketitle

\normalsize

\vspace{-28pt} 
\begin{abstract}
\small
Developing inverse design methods for functional materials with specific properties is critical to advancing fields like renewable energy, catalysis, energy storage, and carbon capture. Generative models based on diffusion principles can directly produce new materials that meet performance constraints, thereby significantly accelerating the material design process. However, existing methods for generating and predicting crystal structures often remain limited by low success rates. In this work, we propose a novel inverse material design generative framework called InvDesFlow-AL, which is based on active learning strategies. This framework can iteratively optimize the material generation process to gradually guide it towards desired performance characteristics. In terms of crystal structure prediction, the InvDesFlow-AL model achieves an RMSE of 0.0423 Å, representing an 32.96$\%$ improvement in performance compared to exsisting generative models. Additionally, InvDesFlow-AL has been successfully validated in the design of low-formation-energy and low-Ehull materials. It can systematically generate materials with progressively lower formation energies while continuously expanding the exploration across diverse chemical spaces. Notably, through DFT structural relaxation validation, we identified 1,598,551 materials with Ehull < 50meV, indicating their thermodynamic stability and atomic forces below 1e-4 eV/Å. These results fully demonstrate the effectiveness of the proposed active learning-driven generative model in accelerating material discovery and inverse design. To further prove the effectiveness of this method, we took the search for BCS superconductors under ambient pressure as an example explored by InvDesFlow-AL. As a result, we successfully identified Li\(_2\)AuH\(_6\) as a conventional BCS superconductor with an ultra-high transition temperature of 140 K. This discovery provides strong empirical support for the application of inverse design in materials science.

\end{abstract}

\vspace{12pt} 

\section*{Introduction}\label{sec1}

The discovery of new materials~\cite{gnome,mattergen} plays a vital role in advancing human production and daily life, such as photovoltaic and battery materials~\cite{Xiao2025},  high-temperature superconducting materials~\cite{Sun2023,Zhou2025}, novel catalysts and degradable materials~\cite{Gao2025}, semiconductor materials~\cite{Jiang2025} and biocompatible materials~\cite{Jiao2025}. The discovery of new materials not only accelerates technological progress but also offers essential solutions to global challenges.
Traditional material discovery faces limitations such as long experimental cycles, uncertain research paths, and a lack of clear exploration directions. While high-throughput screening methods based on density functional theory~(DFT)~\cite{choudhary2022designing} have accelerated the discovery process, they still demand costly computational resources. In recent years, technologies like large language models~\cite{openai2024gpt4technicalreport,crystallm}, geometric graph neural networks~\cite{invdesreview,han2024surveygeometricgraphneural}, and generative AI~\cite{ddpm,song2021Score-based} have demonstrated significant advantages in material generation and screening, further driving new material discoveries. However, current generative models often fail to produce stable materials based on DFT calculations, or are limited to a narrow subset of elements, or can only optimize a very limited set of properties.

For example, generative AI faces limitations in material stability and functionality~\cite{diffcsp,concdvae,mattergen,han2024invdesflowaisearchengine}, where generated materials may fail in practical applications due to thermodynamic instability or inaccurate functional property predictions. Moreover, discriminative AI struggles with overfitting issues, particularly on small datasets, resulting in poor generalization ability across different material systems (e.g., predicting the critical temperature $T_c$ of superconductors~\cite{choudhary2022designing}). Theoretically viable materials may also prove impractical in experiments due to synthesis challenges. Future research needs to address these challenges to shift from generating ``feasible materials'' to producing ``manufacturable materials''.

The current approaches for controlling the generation of functional materials primarily include adapter modules for conditional control~\cite{mattergen} and latent variable decoding of material properties~\cite{concdvae, condcdvae}. Due to the large datasets available for some functional materials, such as formation energy, magnetic density, band gap, or bulk modulus, these generative AI methods perform well. However, for high-temperature superconductors, synthesizable materials, and stable materials, the data is scarce, and the labeling costs are high. In this context, the aforementioned generative AI methods fail to account for these challenges in utilizing minimal data for inverse materials design. 
It is common knowledge that models trained with Adapter Modules do not perform as well as those trained with full-parameter training. Inspired by reinforcement learning, we believe that active learning can maximize model performance.
Active learning's iterative learning cycle allows the model to actively select the most valuable data for improving its performance, significantly enhancing model efficacy, particularly in scenarios where data labeling is expensive or time-consuming.

In this study, we present InvDesFlow-AL, a material inverse design generation framework based on active learning. This framework is capable of directing the generation of target functional materials and continuously optimizing the generation outputs through a step-by-step iterative process. Furthermore, InvDesFlow-AL can adapt to a wide range of downstream tasks by altering the training dataset, thereby achieving efficient inverse material design (as shown in Figure~\ref{main}). To this end, we have combined a generative model that progressively refines atomic types, coordinates, and periodic lattices with an online learning strategy. By selecting more valuable data from the model’s generated results for labeling and training, we have significantly enhanced the performance of the generative model. Compared with existing material generation models, InvDesFlow-AL has reduced the root mean square error (RMSE) in crystal structure prediction tasks to 0.0423 Å, representing a 32.96\% improvement in performance over current methods (see Table~\ref{invdesflow-csp} for details). In terms of generating materials with low formation energy and low Ehull targets, InvDesFlow-AL has successfully identified 1,598,551 materials with an Ehull<50meV (as shown in Figure~\ref{formation-hull}). All these materials have undergone structural relaxation, with interatomic forces less than \(1 \times 10^{-4} \text{eV}/\text{Å}\) (achieving DFT precision). As a proof of concept, InvDesFlow-AL has generated the material with the highest transition temperature in the current conventional superconducting system (Li\(_2\)AuH\(_6\), 140 K) (as shown in Figure~\ref{invdesflow-AL-supercon}). Moreover, this method can also design materials under various property constraints, such as ultra-high-temperature ceramics (as shown in Figure~\ref{uhtc}). Finally, we summarized the current limitations of InvDesFlow-AL in terms of algorithms, data, and scientific theory, and proposed corresponding improvement strategies. We also discussed its potential applications in areas such as electrode materials, hydrogen storage materials, and bioinspired materials, as well as its prospects for facilitating experimental synthesis and industrial deployment.

\section*{Results}\label{sec-results}
\subsection*{Active learning-based diffusion model.}

InvDesFlow-AL is a active learning-based diffusion model for designing target functional inorganic crystal materials across the periodic table.
The diffusion model generate samples by reversing a fixed corruption process. Following the conventions of existing generative models for crystal matrials~\cite{mattergen,diffcsp,cdvae}, we define a crystal material by its unit cell, which includes atom types A (i.e., chemical elements), coordinates X, and periodic lattice L (Supplementary Information section xxx). For each component, we define a corruption process that takes into account its specific geometry and has a physically motivated limiting noise distribution. Unlike previous generative models, we also incorporate an active learning strategy that is actively selecting the most valuable data for labeling and training to enhance its performance. Common strategies for identifying the most valuable data include: the diversity sampling (DS) strategy, which ensures that the selected samples represent different regions of the data distribution; the expected model change (EMC) strategy, which selects samples that have the greatest impact on the model parameters; the query-by-committee (QBC) strategy, which trains multiple models to form a ``committee'' and selects the most valuable samples.

\begin{figure*}[h!]
		\centering  
		\includegraphics[width=1.0\linewidth]{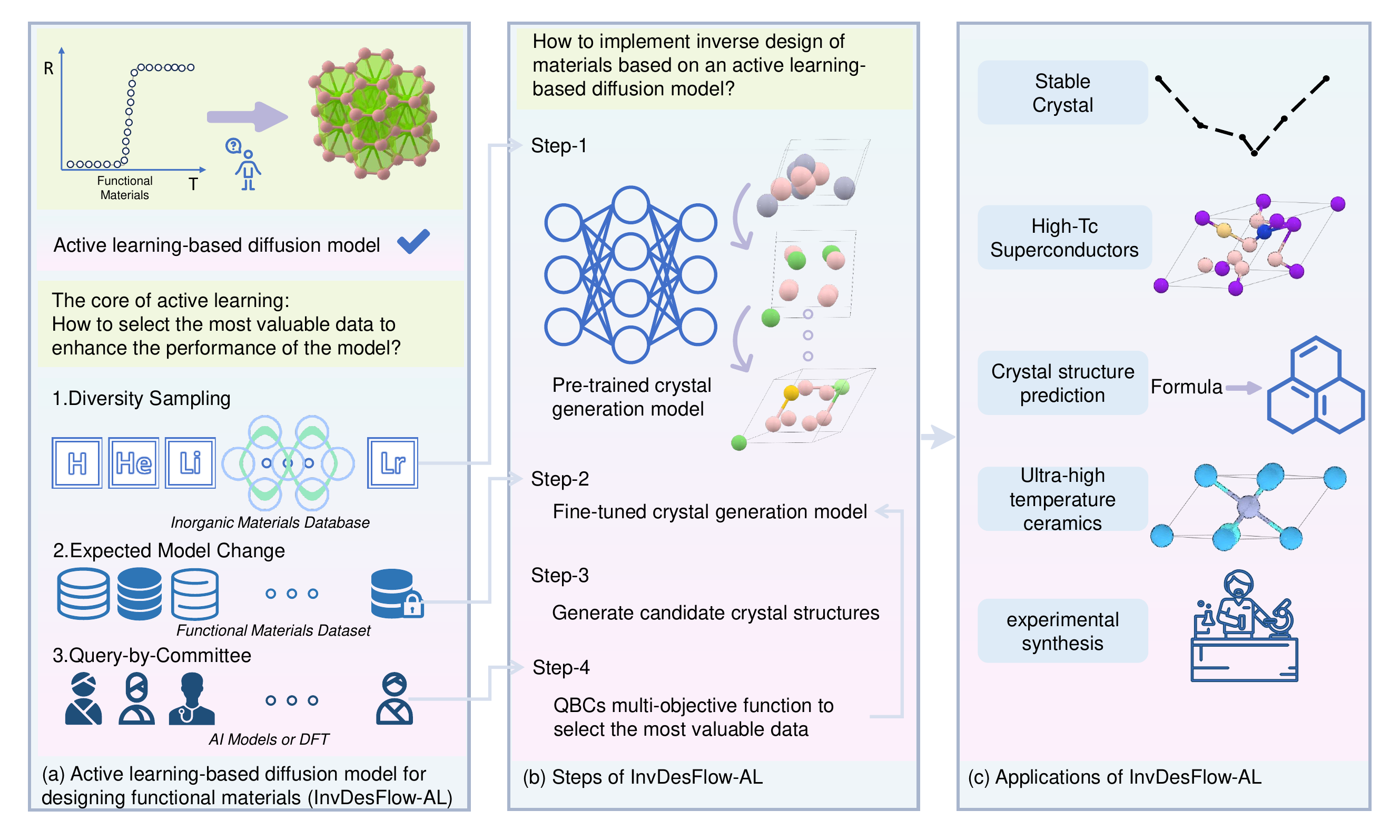}
		\caption{
            \textbf{Active learning-based workflow for inversing design of materials.} (a) Active learning-based diffusion model for designing functional materials. The core of active learning lies in selecting the most valuable data for models to enhance their performance. It primarily involves three strategies: diversity sampling, expected model change, and query-by-committee. (b) Steps of InvDesFlow-AL consist of the following four stages: first, a pre-trained crystal generation model is constructed. Second, this model is fine-tuned on functional materials. Third, the fine-tuned generator is used to generate candidate crystal structures. Finally, a QBCs-based multi-objective function is applied to select the most informative data, which is then used to further fine-tune the generative model. (c) Applications of InvDesFlow-AL. Stable crystal generation, discovery of high-\(T_c\) superconductors, crystal structure prediction, identification of ultra-high temperature ceramics, and guidance for experimental synthesis.
        }
		\label{main} 
\end{figure*}

To train the proposed InvDesFlow-AL, we utilized the Alex-MP-20 dataset~\cite{mattergen}, which was released by MatterGen and contains 607,683 crystalline materials, along with 381,000 inorganic materials~\cite{gnome} from the GNoME dataset.These datasets encompass both ordered and disordered crystal structures and cover a wide range of inorganic materials, including conductors, semiconductors, insulators, and magnetic materials~(see Figure~\ref{main}).
The pretrained model employs an active learning-based diversity sampling strategy to cover different regions of the inorganic materials distribution. It is capable of unconditionally generating crystal structures with high stability, uniqueness, and novelty. Furthermore, the model can be fine-tuned to generate materials with specific functional properties.
InvDesFlow-AL has been successfully applied and demonstrated its feasibility in three tasks: generation of low formation energy materials, generation of high-temperature superconducting materials, and crystal structure prediction. In these tasks, we implemented multiple rounds of fine-tuning (Figure~\ref{main}) for the generative model. The initial fine-tuning utilized crystal data from target functional materials, while subsequent iterations of fine-tuning employed crystal data generated by the model itself. We regard this iterative fine-tuning process as an implementation of the expected model change strategy.
For different tasks, InvDesFlow-AL using different QBC methods to select suitable crystal data. These QBCs are designed as multi-objective functions and are used to evaluate the most valuable data.

The pre-trained crystal generation mode of InvDesFlow-AL is trained on a large-scale inorganic materials dataset, demonstrating strong performance in material novelty (Supplementary Fig. S4) and excellent stability (Figure~\ref{formation-hull}). Compared to MatterGen~\cite{mattergen}, which adopts adapter-based fine-tuning for inverse design of functional materials, InvDesFlow-AL leverages full-parameter fine-tuning across the entire architecture. This approach demonstrates better adaptability to downstream tasks, as extensively validated in domains such as large language models~\cite{lora,adapter-nlp} and computer vision~\cite{adapter-cv}.
For crystal representation, InvDesFlow-AL adopts fractional coordinates anchored to lattice vectors as the fundamental basis, which explicitly preserves crystalline periodicity and enables more intuitive handling of crystal symmetry. This methodology ensures higher efficiency and stability during both model training and sampling processes, a conclusion further substantiated by subsequent crystal structure prediction tasks.

\subsection*{Generation of low formation energy materials}
The pursuit of synthesizing materials with low formation energy~(\(E_{\text{form}}\)) and minimal energy above the convex hull (\(E_{\text{hull}}\)) constitutes the central objective of InvDesFlow-AL. The formation energy, quantifying the thermodynamic stability of a compound relative to its elemental constituents, serves as a fundamental indicator of synthesizability—only materials with negative \(E_{\text{form}}\) are thermodynamically viable under equilibrium conditions. Energy above hull (\(E_{\text{hull}}\)), conversely, measures a material's metastability by evaluating its energetic proximity to the convex hull in phase space. A low \(E_{\text{hull}}\) (typically < 50 meV/atom) signifies resilience against decomposition into competing phases, a critical prerequisite for practical synthesis and operational durability.

In this section, we focus on the ability of the InvDesFlow-AL to generate materials with low formation energy (\(E_{\text{form}}\)) and minimal energy above the convex hull (\(E_{\text{hull}}\)). To achieve thermodynamically favorable (low \(E_{\text{form}}\)), structurally stable (interatomic forces < 1e-4 eV/Å), and compositionally novel candidates, we propose an active learning framework integrating EMC and QBC strategies.  
\begin{figure*}[h!]
		\centering  
		\includegraphics[width=1.0\linewidth]{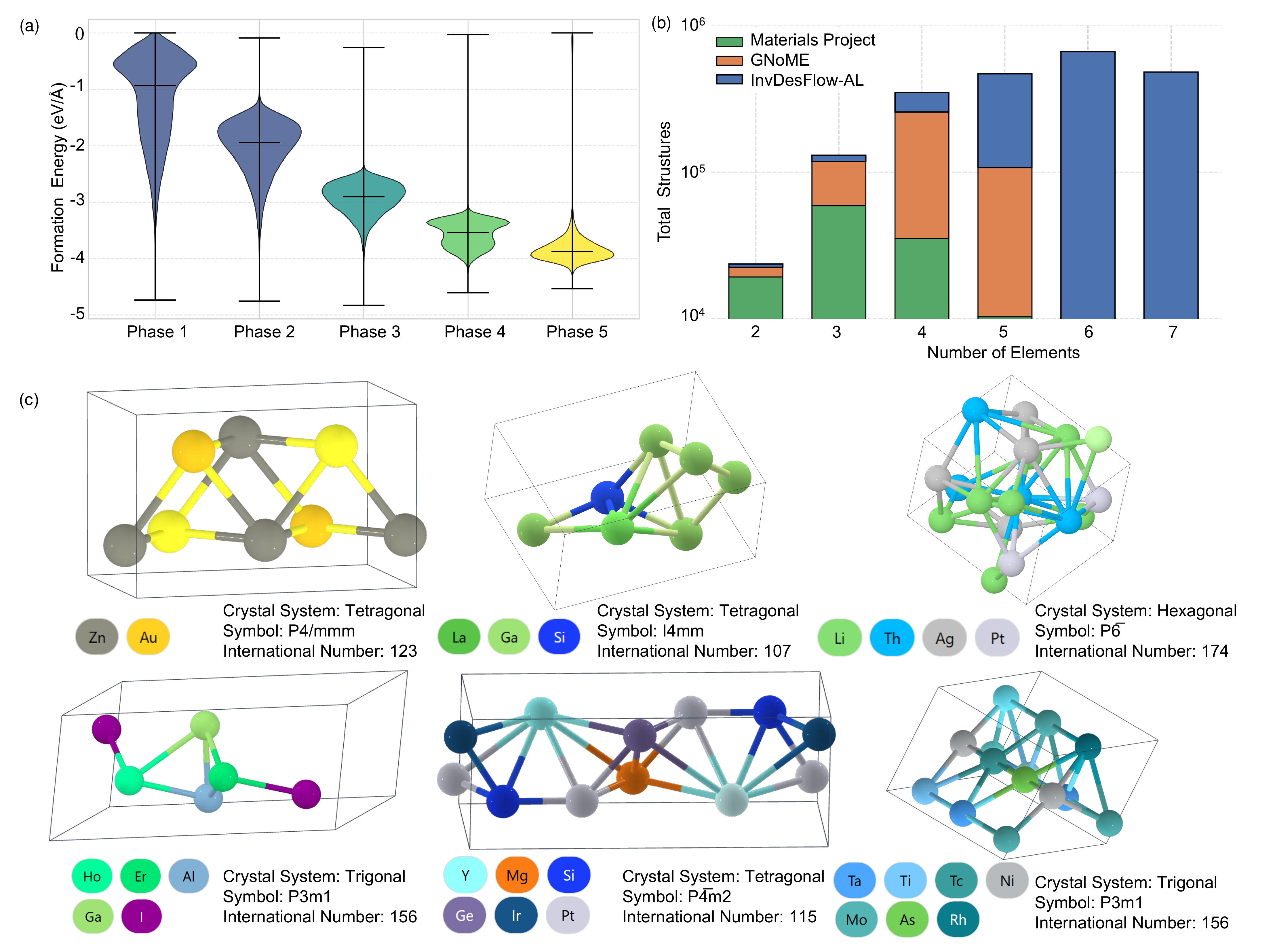}
		\caption{
            \textbf{InvDesFlow-AL for the generation of low formation energy materials.}  
(a) InvDesFlow-AL employs a crystal generation model to generate new crystal structures, followed by formation energy prediction using FormEGNN. A lower threshold is applied to retain newly generated materials for fine-tuning the generative model. After five iterations, a progressive decrease in formation energy is observed.  
(b) InvDesFlow-AL adopts the same strategy to generate materials with low E$_\text{hull}$. Through multiple rounds of generation, a total of 1,610,600 new crystal structures with E$_\text{hull}$ < 50 meV have been obtained, expanding the chemical space exploration to a broader range of atomic species.  
(c) Crystal structures containing 2, 3, ..., and up to 7 elements generated by InvDesFlow-AL.
        }
		\label{formation-hull} 
\end{figure*}

The pretrained model is fine-tuned on the GNoME dataset~\cite{gnome}, focusing exclusively on crystals with \(E_{\text{form}} < -0.5\) eV to establish thermodynamic stability priors. The fine-tuned generator synthesizes novel crystal structures, filtered by compositional uniqueness against existing materials databases~(e.g., Materials Project~\cite{materials-project}). Generated candidates undergo atomic-scale structural relaxation using the DPA-2 interatomic potential~\cite{dpa2}, which achieves DFT-level accuracy. Structures failing force convergence criteria (\(||\mathbf{F}|| > 1e-4\) eV/Å) are systematically discarded. The formation energy prediction model~(which we introduce here as FormEGNN) was developed in InvDesFlow-1.0~\cite{InvDesFlow}. It predicts \(E_{\text{form}}\) for relaxed structures, with a dynamically adjusted threshold to retain only the most stable candidates for subsequent generator retraining. The iterative refinement process is driven by EMC-guided data selection, where a committee consisting of DPA-2 and FormEGNN evaluates candidate materials using a multi-objective scoring function. Section~\ref{sec-method} provides a detailed explanation of this multi-objective function. This closed-loop paradigm progressively biases the generator toward materials with low formation energy and relaxed structures while maintaining structural diversity.

InvDesFlow-AL iteratively generated and fine-tuned the model five times. The average formation energies of the generated crystals in these five iterations were \(\mu = -1.14, -2.03, -2.93, -3.56, -3.77\), with the number of generated structures being 80,707, 95,580, 97,379, 136,784, and 166,663, respectively, resulting in a total of 577,113 generated crystals. The dataset has been open-sourced~\footnote{https://zenodo.org/records/15222702}. All these structures achieved DFT relaxation accuracy. As shown in Figure~\ref{formation-hull} (a), the formation energy of the generated crystals decreases with increasing iterations.

InvDesFlow-AL generates low-hull crystal structures through iterative optimization. During each iteration, the Bohrium platform~\footnote{https://github.com/deepmodeling/openlam} is utilized to query the E$_\text{hull}$ values of materials, and crystals with lower E$_\text{hull}$ values are selected to fine-tune the model. We have performed 10 rounds of fine-tuning in total, generating 1,598,551 materials with formation energies below 50 meV (see Supplementary Fig. S5). As shown in Figure~\ref{formation-hull} (b), to generate lower-E$_\text{hull}$ crystal structures, the elemental composition distribution of the generated crystals by InvDesFlow-AL differs significantly from those in the Materials Project~\footnote{https://legacy.materialsproject.org/} database. InvDesFlow-AL exhibits a strong preference for generating materials with higher atomic diversity, many of which are high-entropy alloys that do not exist in existing databases. While binary and ternary chemical spaces will likely be exhaustively explored by chemists in the near future, despite the experimental challenges in synthesizing multi-component materials, we believe that these complex multi-element crystals will make groundbreaking contributions to future advanced materials discovery, Figure~\ref{formation-hull} (c) shows the multicomponent crystal structures generated by InvDesFlow-AL. All generated crystal structures will be fully open-sourced to the community~\footnote{https://zenodo.org/records/15221067}.

\subsection*{Generation of high-temperature superconducting materials} 
In this section, we apply InvDesFlow-AL to the discovery of high-temperature superconducting materials. High-temperature superconductors hold significant importance for achieving efficient power transmission, controlled nuclear fusion energy generation, magnetic resonance imaging, and superconducting quantum computing. Superconducting materials surpassing critical temperature thresholds - including the McMillan limit (40 K), liquid nitrogen temperature regime (77 K), and even room temperature - attract distinct levels of interest from physicists. For instance, hydrogen-based systems like H$_3$S (200 K)~\cite{Drozdov2015} and LaH$_{10}$ (260 K)~\cite{Drozdov2019,LaH10Tc260} with exceptionally high transition temperatures have drawn extensive research attention. Subsequent discoveries reported in~\cite{Zhong2022}, including YH\textsubscript{18} (183 K), AcH\textsubscript{18} (206 K), LaH\textsubscript{18} (271 K), and ThH\textsubscript{18} (306 K), have further expanded the family of hydride-based superconducting materials. However, these materials require extreme high-pressure conditions for synthesis, which poses significant constraints on their practical applicability. Consequently, the discovery of high-temperature superconductors under ambient pressure has become particularly crucial. Notably, the experimental synthesis~\cite{Zhou2025} of (La,Pr)$_3$Ni$_2$O$_7$ (45 K) under ambient pressure has garnered considerable attention as it exceeds the McMillan limit. Recent advancements in high-throughput computing and machine learning techniques have accelerated the discovery of superconducting materials, with Mg$_2$XH$_6$ (X = Rh, Ir, Pd, or Pt)~\cite{Sanna2024, PhysRevLett.132.166001,AFM202404043,ZHENG2024101374}  exhibiting a T$_{c}$ exceeding 80 K. These developments motivate our adoption of state-of-the-art AI technologies to further expedite the exploration of superconducting materials.

The conventional high-throughput screening approach, constrained by the limited exploration of chemical space, has shown fundamental limitations in discovering high-temperature superconductors. Our InvDesFlow-AL framework enables direct generation of materials with targeted properties, facilitating exploration in unbounded chemical space. To generate materials with ambient-pressure high-temperature superconductivity, we develop a multi-stage active learning strategy combining sequential model refinement and committee-based validation.
The strategy initiates with a two-phase expected model change optimization. Phase I conducts domain adaptation through fine-tuning on metallic materials from the pre-training dataset, establishing essential electronic conductivity priors. Phase II implements superconducting specialization using recently discovered conventional superconductors~\cite{Sanna2024, PhysRevLett.132.166001, AFM202404043, ZHENG2024101374}, progressively aligning the model's generative space with superconducting characteristics. This sequential refinement ensures structural validity while enhancing target property awareness.

The optimized generator produces candidate superconductors that undergo dual-stage validation. We firstly introduce an superconducting graph neural network~(SuperconGNN) to screen candidates by predicted critical temperature (T$_{c} > $  20 K threshold). Subsequently, we adopt DFT calculations to verify electronic structure features and superconducting stability. Validated materials enrich the training dataset through an active learning loop governed by a query-by-committee strategy. The iterative refinement process is driven by QBC-guided data selection, where a committee consisting of SuperconGNN and DFT evaluates candidate materials using a multi-objective scoring function. Section~\ref{sec-method} provides a detailed explanation of this multi-objective function.

\begin{figure*}[h!]
		\centering  
		\includegraphics[width=1.0\linewidth]{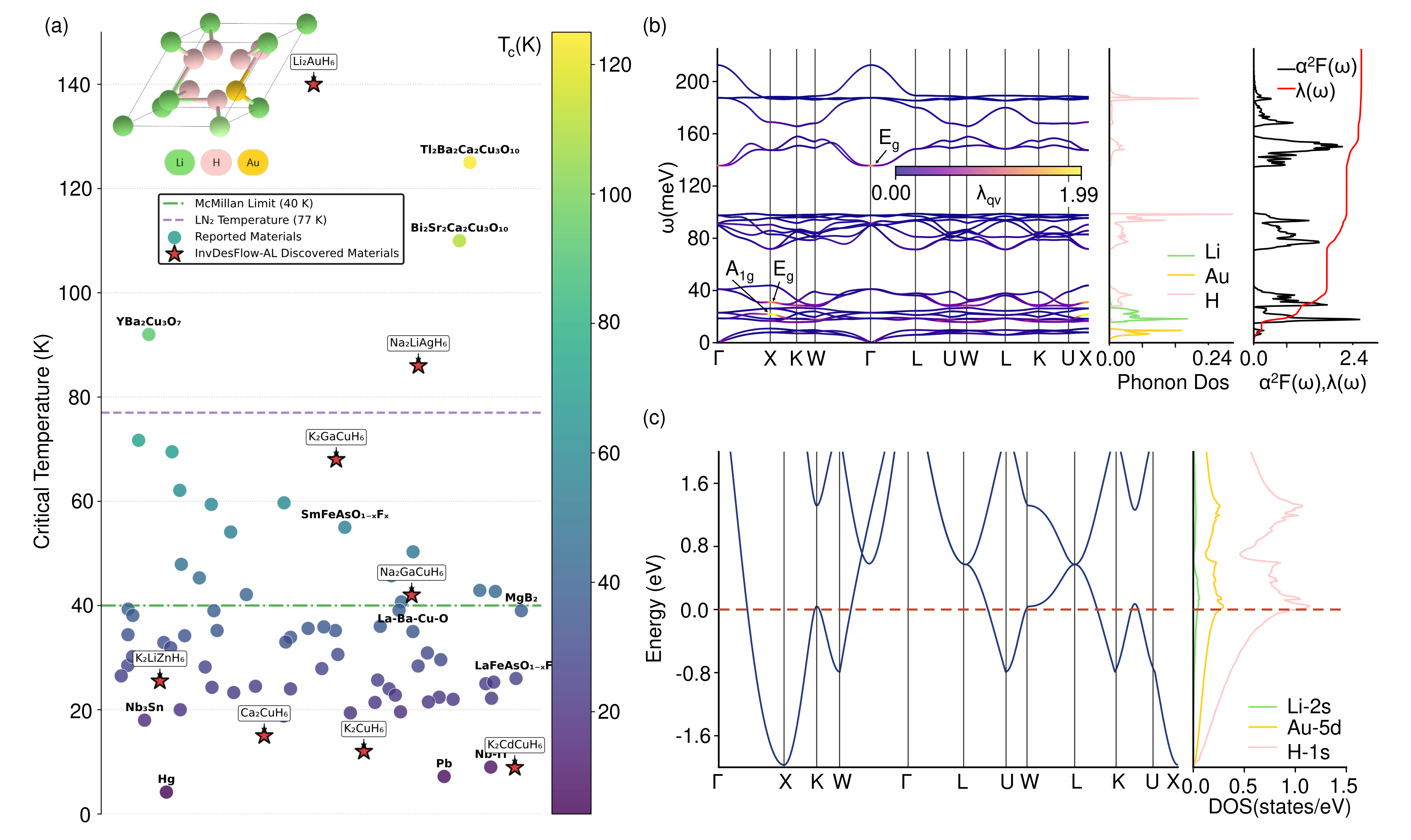}    
		\caption{
\textbf{InvDesFlow-AL for discovering novel high-temperature superconducting materials. }
(a) Comparison of previously reported high-temperature superconductors and newly discovered materials generated by InvDesFlow-AL. The superconducting transition temperatures of these newly discovered materials span a wide range, from the McMillan limit to the liquid nitrogen temperature region. The inset in (a) shows the crystal structure of Li$_2$AuH$_6$. (b) Phonon dispersion and phonon density of states of Li$_2$AuH$_6$, indicating its dynamical stability. (c) Electronic band structure and density of states of Li$_2$AuH$_6$.
        }
		\label{invdesflow-AL-supercon} 
\end{figure*}

As shown in Figure~\ref{invdesflow-AL-supercon} (a), through multiple iterative generation cycles, InvDesFlow-AL has identified superconducting materials spanning a wide temperature range, from low to high T$_c$. Notably, K$_2$GaCuH$_6$ and Na$_2$GaCuH$_6$ exhibit critical temperatures exceeding the McMillan limit (40 K), while Li$_2$AuH$_6$ and Na$_2$LiAgH$_6$ surpass the liquid nitrogen temperature threshold (77 K). Remarkably, Li$_2$AuH$_6$ reaches approximately 140 K, representing the highest T$_c$ achieved by conventional superconductors to date.

Figure~\ref{invdesflow-AL-supercon} (b) presents the phonon dispersion relations and phonon density of states (DOS) of Li$_2$AuH$_6$ , demonstrating its dynamical stability. 
Figure~\ref{invdesflow-AL-supercon} (c) presents the electronic structure of Li$_2$AuH$_6$, revealing its metallic character with multiple bands crossing the Fermi level. The atomic orbital-projected density of states (PDOS) analysis demonstrates that electronic states near the Fermi level are predominantly contributed by the Au-H octahedral coordination. A van Hove singularity is identified at the W point, generating a pronounced PDOS peak at the Fermi level, which is often associated with enhanced electronic correlations. The phonon dispersion of Li$_2$AuH$_6$ exhibits no imaginary frequencies at ambient pressure, confirming its dynamic stability. The spectrum consists of three distinct frequency regions separated by two energy gaps: the high-frequency region (\(>120\) meV), dominated by hydrogen vibrations; the intermediate-frequency region (\(60-100\) meV), primarily composed of H-related vibrations with partial Au contributions; and the low-frequency region (\(<50\) meV), where mixed vibrations of Au and Li atoms are observed. The Eliashberg spectral function \(\alpha^2F(\omega)\) and the cumulative electron-phonon coupling (EPC) constant \(\lambda(\omega)\) are displayed in the inset, with the total EPC constant calculated to be \(\lambda = 2.84\). The strong EPC predominantly arises from three key phonon modes: the \(E_g\) mode at \(\Gamma\) (140 meV), the \(A_{1g}\) mode at X (20 meV), and the \(E_g\) mode at X (30 meV). First-principles analysis reveals that the robust electron-phonon interactions in Li$_2$AuH$_6$ originate primarily from the vibrational modes of the Au-H octahedral framework and Li atoms, which induce significant charge density modulations and drive strong Cooper pairing. This exceptionally large \(\lambda\) value places Li$_2$AuH$_6$ among the strongest electron-phonon coupled superconductors predicted to date, highlighting its potential as a promising high-temperature conventional superconductor. 
For a more detailed analysis of the Li$_2$AuH$_6$ crystalline material and its synthesis pathway, please refer to our another work~\cite{liauh}. A variety of additional superconducting materials were also discovered by InvDesFlow-AL, with their corresponding DFT calculation results presented in Supplementary Note A.

\subsection*{Stable structure prediction task}
\label{csp-task}

The crystal structure prediction task (CSP) is of great significance for accelerating the discovery of new materials, understanding the fundamental laws of matter, and guiding experimental synthesis. In recent years, AI methods such as CDVAE~\cite{cdvae}, DiffCSP~\cite{diffcsp}, and CrystaLLM~\cite{crystallm} have achieved promising results in this task. However, these methods often require multiple samplings and do not provide a clear approach for selecting the best structure from the multiple predictions. As a result, the results from multiple samplings do not necessarily reflect the true prediction accuracy. We have significantly improved the accuracy of crystal structure prediction using an active learning-based approach, with the single-prediction accuracy surpassing the current best methods.

It is important to note that the functional material generation described in previous sections employs a de novo generative model , where the number of atoms in the unit cell serves as a conditional constraint. In contrast, the CSP task focuses on predicting crystal structures with fixed atomic compositions. For the CSP implementation, we first adopted a diversity sampling strategy to ensure the model learns structural information across diverse data distributions. This process utilized the Alex-MP-20 dataset released by MattergGen~\cite{mattergen} and crystal structures from GNoME~\cite{gnome}, with rigorous exclusion of test sets including Perov-5, MP-20, and MPTS-52 during data preprocessing.
Using the expected model change strategy, we fine-tuned the model on the corresponding structure prediction training sets (Perov-5, MP-20, MPTS-52).

Additionally, we utilized de novo generative model and fine-tuned it on the aforementioned training set. Based on the atomic number distribution, we generated 100,000 crystal structures. For the generated materials, structural relaxation was performed using the DPA2 potential function to filter out structurally stable materials. Subsequently, we employed the formation energy prediction model FormEGNN to identify crystal structures with the same chemical formula but lower structural energy, and then fine-tuned the model again. The feedback from DPA2 and FormEGNN is referred to as an active learning-based query-by-committee, where the committee assists in selecting the most valuable samples.
A multi-objective scoring function is used here, which will be introduced in section~\ref{sec-method}.

As shown in Table~\ref{invdesflow-csp}, MP-20 is a test set with a maximum of 20 atoms per unit cell, representing conventional crystalline materials in existing databases like the Materials Project~\footnote{https://legacy.materialsproject.org/}~\cite{materials-project}. On the MP-20 test set, InvDesFlow-AL achieved an RMSE of 0.0423 Å, surpassing not only the recently popular large language model-based crystal structure prediction algorithm CrystaLLM~\cite{crystallm} but also outperforming state-of-the-art methods such as DiffCSP~\cite{diffcsp}, CDVAE~\cite{cdvae}, and EquiCSP~\cite{equicsp}. MPTS-52, a dataset allowing up to 52 atoms per unit cell, further demonstrated the robustness of our approach. InvDesFlow-AL achieved an RMSE of 0.0725 Å, representing a 37$\%$ improvement over EquiCSP~\cite{equicsp}.

\begin{table*}[h!]
  \centering
  \caption{Results on crystal structure prediction task. MR stands for Match Rate.}
  \resizebox{\linewidth}{!}{
  \small
  \setlength{\tabcolsep}{3.2mm}
    \begin{tabular}{lcccccc}
    \toprule
     &  \multicolumn{2}{c}{Perov-5 } & \multicolumn{2}{c}{MP-20} & \multicolumn{2}{c}{MPTS-52} \\
         & MR (\%) & RMSE & MR (\%) & RMSE  & MR (\%) & RMSE \\
    \midrule
    RS & 36.56  & 0.0886  & 11.49  & 0.2822 & 2.68 &	0.3444  \\
    BO & 55.09  & 0.2037  & 12.68  & 0.2816 & 6.69 &	0.3444 \\
    PSO & 21.88  & 0.0844  & 4.35  & 0.1670 & 1.09 &	0.2390  \\
    \midrule
     P-cG-SchNet~\cite{gebauer2022inverse}    & 48.22  & 0.4179   & 15.39 & 0.3762 & 3.67 & 0.4115  \\   
    CDVAE~\cite{cdvae}   & 45.31  & 0.1138  & 33.90  & 0.1045 & 5.34 & 0.2106 \\
    DiffCSP~\cite{diffcsp}      & 52.02  & 0.0760  & 51.49 &	0.0631 & 12.19 & 0.1786 \\
    CrystaLLM~\cite{crystallm}      & 47.95  & 0.0966  & 55.85 &	0.0437 & 17.47 & 0.1113 \\
    EquiCSP~\cite{equicsp} & 52.02 &  0.0707 & 57.39 & 0.0510 & 14.85 & 0.1169 \\
    \midrule

    InvDesFlow-AL & \textbf{52.86} & \textbf{0.0703} & \textbf{60.83} & \textbf{0.0423} & \textbf{23.72} & \textbf{0.0725} \\
    
    \bottomrule
    \end{tabular}%
  \label{invdesflow-csp}%
  }
  \vspace{-2ex}
\end{table*}

\subsection*{Other Application Demonstrations}
Ultra-high temperature ceramics (UHTCs)~\cite{Wyatt2024} constitute a category of advanced materials that preserve exceptional mechanical robustness and thermophysical stability under extreme conditions, including ultra-high temperatures (>2000 °C), aggressive oxidation, and corrosive atmospheres. These materials exhibit critical application value in aerospace, defense, and energy engineering. UHTCs are predominantly composed of refractory transition-metal borides, carbides, and nitrides, such as borides (XB$_2$, X = Zr, Hf, Ta) and carbides (XC, X = Zr, Hf, Ta, Ti), all of which possess melting points exceeding 3000 °C. Owing to their superior integrated properties—encompassing oxidation/ablation resistance, high-temperature strength retention, and thermal shock resilience—UHTCs have attracted extensive scientific attention, particularly for applications in thermal protection systems of hypersonic vehicles and next-generation nuclear reactors.

\begin{figure*}[h!]
		\centering  
		\includegraphics[width=0.9\linewidth]{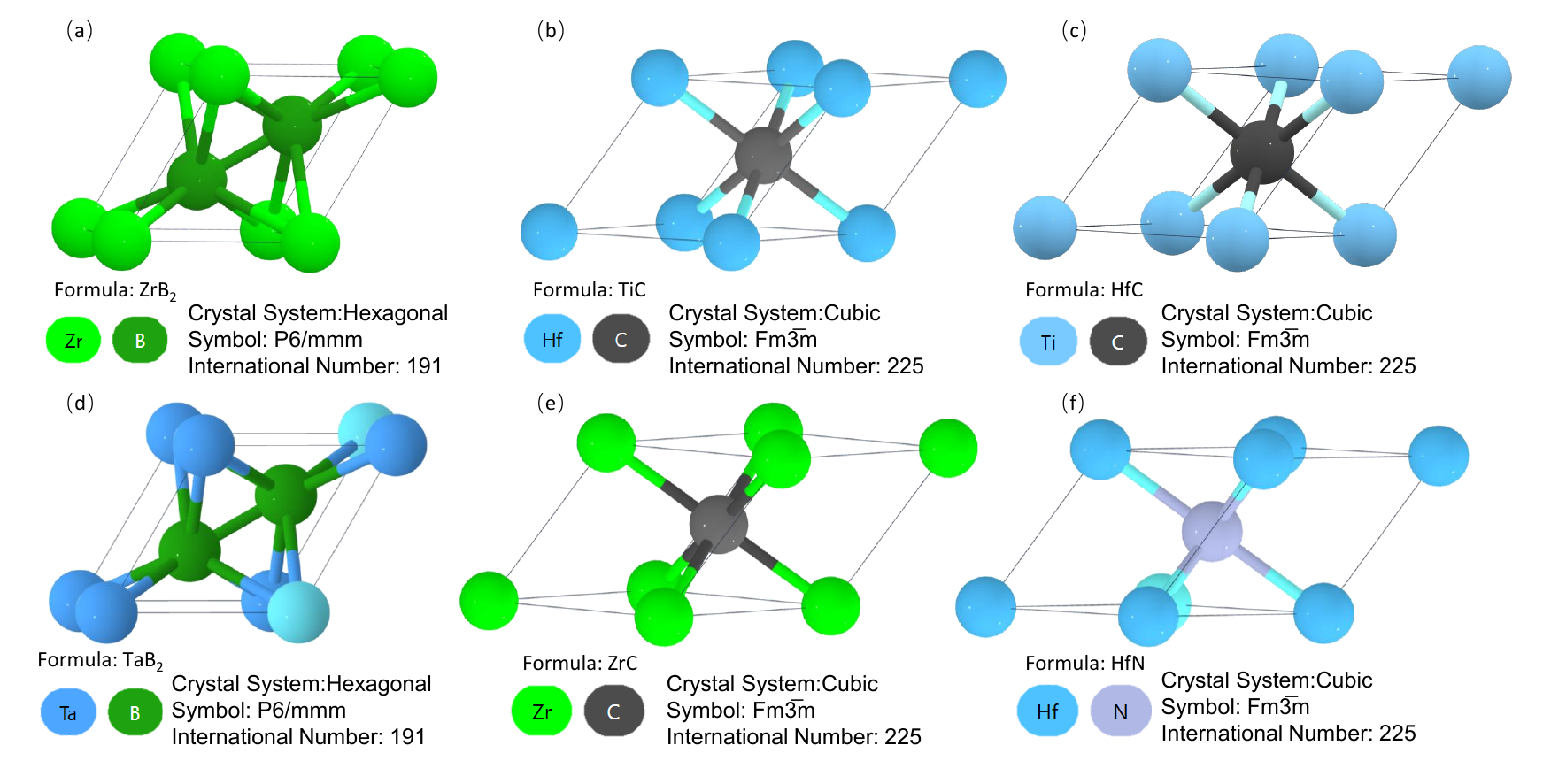}
		\caption{
        \textbf{InvDesFlow-AL for generating ultra-high-temperature ceramics.}  
(a)-(c) Synthesized UHTCs (ZrB$_2$, HfC, TiC) existing in the fine-tuning dataset, demonstrating the model’s capacity to reproduce known high-performance ceramics.  
(d)-(f) Novel UHTCs (TaB$_2$, ZrC, HfN) absent from the training data, whose exceptional properties—high-temperature resistance, oxidation resistance, and thermal shock resilience—have been validated through theoretical/experimental studies. This systematic validation underscores InvDesFlow-AL’s capability to design unreported, high-stability ceramics beyond existing databases. 
        }
		\label{uhtc} 
\end{figure*}

We applied InvDesFlow-AL to generate UHTCs. This section aims to demonstrate the generalization capability of the pretrained model. We collected crystal structure data for 14 UHTCs, such as ZrB$_2$, HfC, TiC. Based on the pre-trained crystal generation model, we first fine-tuned it on the 14 UHTC crystal materials. Using the fine-tuned generative model, we then generated 100 crystal structures, achieving 100$\%$ coverage of the 14 UHTC materials in the training set. Among the remaining generated materials, we checked whether their chemical formulas had been previously synthesized or calculated using DFT. Three materials (Figure~\ref{uhtc} (d)-(f)), TaB$_2$, ZrC, and HfN, were found to have been experimentally validated. TaB$_2$ has a high melting point and excellent wear resistance, making it an ideal choice for ultra-high-temperature ceramic materials~\cite{uhtc-TaB2}. It can be synthesized using high-temperature synthesis~\cite{uhtc-TaB2-HighTem}, chemical vapor deposition, and sol-gel techniques. ZrC has a NaCl-type structure with a melting point of 3540°C and is used as a nuclear fuel cladding material~\cite{uhtc-ZrC}. HfN exhibits high infrared reflectance and can be applied as a thermal control coating for satellites~\cite{uhtc-HfN}.

\section*{Discussion}\label{sec12}

From electronic devices and aerospace to quantum computing and energy transport and storage, the advancement of almost every frontier technology depends on breakthroughs in novel functional materials. We introduce InvDesFlow-AL, an active learning-based inverse materials design framework. Through InvDesFlow-AL, scientists can specify target properties—such as superconductivity, topology, and magnetism—to generate candinate crystal structures, thereby overcoming the limitations of empirical approaches. InvDesFlow-AL integrates generative AI, quantum mechanical methods, and thermodynamic and kinetic modeling to enable multiscale simulation for the discovery of new functional materials. In crystal-structure prediction, our model  achieves a RMSE of 0.0423 Å, which is close to the typical systematic error observed in X-ray diffraction (XRD) measurements due to instrument calibration, sample preparation, and data processing. In the generation of low-formation-energy and low-Ehull materials, InvDesFlow-AL demonstrates the effectiveness of active learning through multiple iterations. Each round produces materials with progressively lower formation energies, ultimately generating 1,598,551 materials with Ehull < 50 meV. For high-temperature superconductor discovery, our framework identified Li\(_2\)AuH\(_6\) as a candidate with an ultra-high superconducting transition temperature of 140 K. Additionally, we discovered a series of novel superconductors that span the range between the McMillan limit and the liquid nitrogen temperature regime.
Outstanding materials of interest to humanity often lie outside existing crystal structure databases, limiting the performance of adapter-based or latent-space-controlled generative models when the training data is insufficient. InvDesFlow-AL is designed to overcome this constraint. By leveraging an active learning strategy, the framework uses data generated by the model, filters it using QBCs, and fine-tunes the generative model over multiple iterations. As a result, InvDesFlow-AL continuously evolves to generate increasingly superior functional materials.

However, whether the generative model within InvDesFlow-AL can successfully discover superior materials depends critically on the quality of the pretrain data, architecture of the generative model, the design of the QBCs strategy, and the advancement of scientific theory. Different classes of functional materials exhibit varying degrees of data availability. For instance, high-temperature superconductors under ambient pressure are extremely scarce, whereas more than 20,000 superconductors with transition temperatures below 10 K have already been identified. Traditional high-throughput DFT calculations provide a broader dataset, and with the aid of machine learning, this field is expected to yield an increasing number of high-quality data points~\cite{SCHMIDT2024101560}. 
We are currently using an equivariant GNN architecture based on EGNN~\cite{egnn} to implement a crystal generation model. However, AlphaFold 3~\cite{alphafold3} demonstrates that equivariance can be achieved through data augmentation (random rotations and translations), effectively eliminating the need for architectural equivariance, while achieving state-of-the-art performance in molecular docking. This inspires us to explore crystal representation architectures beyond the constraints of equivariant networks. We currently use a QBC framework composed of AI models and DFT calculations to effectively screen for high-quality data. However, the computational cost of DFT remains prohibitive. In the future, AI models could fully replace DFT, enabling the screening of hundreds of millions of hypothetical crystal structures. Tools such as DeepH~\cite{Li2023}, which aim to replace first-principles calculations with AI, are a promising step in this direction, although they are currently unable to perform fully end-to-end predictions for arbitrary crystal structures. Furthermore, the development of scientific theory itself may impose limitations on new materials discovery. For example, recent studies~\cite{MaximumTc} have suggested that Li\(_2\)AuH\(_6\) and Li\(_2\)AgH\(_6\) represent the upper bound of the superconducting transition temperature for conventional BCS superconductors under ambient pressure. If this conclusion holds true, it would imply that the search for new materials within this system is nearly exhausted. 

With the deep integration of artificial intelligence into materials science, we anticipate that the generalized InvDesFlow-AL framework will continue to unleash its potential in several key materials domains in the future.
In the domain of high-performance battery electrode materials, enhancing energy density fundamentally relies on the development of electrodes with both high capacity and high voltage~\cite{Wang2025}. The InvDesFlow-AL framework can construct multi-objective scoring function—based on theoretical capacity, voltage window, and electrochemical stability, thereby enabling the generation of novel anode and cathode candidates that simultaneously exhibit high theoretical capacity and high operating voltage~\cite{cgcnn-screen}. 
In the context of the green energy transition, the development of lightweight, high-capacity hydrogen storage materials~\cite{Mg-H-review} remains a critical bottleneck for the widespread adoption of hydrogen energy. Conventional approaches struggle to systematically identify materials that simultaneously meet key criteria such as low dehydrogenation temperature, high reversible capacity, and appropriate thermodynamic stability. The InvDesFlow-AL framework offers a natural advantage in optimizing hydrogen storage performance by constructing multi-objective scoring functions targeting hydrogen adsorption/desorption characteristics~\cite{MgH-nc}. This enables the discovery of novel magnesium-based alloys, hydrides, and even composite materials with enhanced hydrogen storage capabilities.
For next-generation bioinspired robotics, mechanical flexibility and stimuli responsiveness are essential for achieving closed-loop control encompassing perception, response, and regulation~\cite{Hao2022}. Looking ahead, the InvDesFlow-AL framework can be extended by constructing multi-objective scoring functions targeting properties such as flexibility, shape-memory behavior, and self-healing capabilities, thereby facilitating the rational design of novel intelligent materials for bioinspired applications. This approach has the potential to significantly accelerate the development of advanced humanoid robots.
The ultimate goal of InvDesFlow-AL is to bridge computational discovery with experimental synthesis and industrial application, going beyond theoretical predictions. Leveraging its active learning mechanism, the candidate pool can be dynamically refined to prioritize the synthesis of high-value materials. Moreover, by integrating metrics such as cost evaluation and environmental sustainability, the framework offers robust decision-making support for real-world applications. By continuously expanding the functional boundaries and application depth of InvDesFlow-AL, we anticipate significant advancements in multiple strategic domains, driving transformative breakthroughs in next-generation energy, intelligent manufacturing, and bioengineering.

\section*{Methods}
\label{sec-method}
\subsection*{InvDesFlow-AL pretrained crystal generation model}

Generating target functional materials is a crucial step in the inverse design of materials. However, prior to achieving desired attributes, generative models must ensure that synthesized materials inherently exhibit fundamental crystalline characteristics, including periodicity, symmetry, interatomic interactions, and chemically reasonable stoichiometry. To this end, we propose a pretrained crystal generation model. In the following, we will introduce the data representation, model architecture, and training method required for this model.

In crystalline materials, atoms arrange themselves in a periodic configuration where the fundamental building block is termed the unit cell, denoted as $\gM = (\mA, \mF, \mL)$. This structural unit consists of three primary components: $\mA = [\va_1, \va_2, ..., \va_N] \in \sR^{h \times N}$ encodes the chemical species present in the unit cell, with h representing the dimensionality of atomic feature descriptors. The spatial arrangement of atoms is specified by $\mF = [\vf_1, \vf_2, ..., \vf_N] \in \sR^{3 \times N}$, which records their three-dimensional fractional coordinates. The periodicity of the crystal framework is defined through the lattice matrix $\mL = [\vl_1, \vl_2, \vl_3] \in \sR^{3 \times 3}$, whose column vectors establish the basis vectors spanning the crystalline lattice system.
This pretrained model adopts the equivariant graph neural network (EGNN) architecture~\cite{egnn,diffcsp}, which ensures translation equivariance by using relative coordinate differences and achieves rotation/reflection equivariance through squared relative distances. Additionally, EGNN maintains the permutation equivariance of graph neural networks, ensuring that permuted input atom orders yield correspondingly permuted outputs. It is worth noting that there are many frameworks for achieving equivariance, such as e3nn~\cite{diffdock}~\footnote{https://e3nn.org/} based on spherical harmonic representations, tensor field networks~\cite{tfn}, and other high-order tensor product networks. These network architectures can be easily integrated into our active learning-based training.
The pretrained model is based on a diffusion generative model. During the generation process, the number of atoms remains unchanged, while atomic types and the lattice matrix undergo noise addition and denoising using the standard enoising diffusion probabilistic model~\cite{ddpm} framework. Fractional coordinates, on the other hand, are processed using a score-matching-based framework~\cite{song2021Score-based}.
The loss function of the pretrained and fine-tuned crystal generation model is given by:  
$\mathcal{L}_{\text{total}} = \mathcal{L}_{\text{lattice}} + \mathcal{L}_{\text{atom}} + \mathcal{L}_{\text{coord}}$. The detailed training and sampling processes are shown in Algorithm~\ref{alg:train_v2} and Algorithm~\ref{alg:gen_v2}.
Training details and hyperparameter settings are provided in Supplementary Note B.
\begin{algorithm}[H]
\small
\caption{Training Procedure for InvDesFlow-AL Pretrained Crystal Generation Model}\label{alg:train_v2}
\begin{algorithmic}[1]
\REQUIRE Crystal structure ($\mL_0$, $\mA_0$, $\mF_0$), atom count $N$, denoising network $\phi$, diffusion steps $T$
\ENSURE Optimized crystal generation model
\STATE Sample timestep $t \sim \mathcal{U}\{1,T\}$ and noise vectors:
\STATE \quad $\vepsilon_\mL,\vepsilon_\mA,\vepsilon_\mF \sim \mathcal{N}(\mathbf{0},\mathbf{I})$
\STATE Compute noise scheduling coefficients $\sqrt{\bar{\alpha}_t}$, $\sqrt{1-\bar{\alpha}_t}$, and $\sigma_t$
\STATE Generate perturbed crystal components:
\STATE \quad $\mL_t = \sqrt{\bar{\alpha}_t}\mL_0 + \sqrt{1-\bar{\alpha}_t}\vepsilon_\mL$ \COMMENT{Lattice diffusion}
\STATE \quad $\mA_t = \sqrt{\bar{\alpha}_t}\mA_0 + \sqrt{1-\bar{\alpha}_t}\vepsilon_\mA$ \COMMENT{Atom type diffusion}
\STATE \quad $\mF_t = w(\mF_0 + \sigma_t\vepsilon_\mF)$ \COMMENT{Periodic coordinate perturbation}
\STATE Predict noise components via network:
\STATE \quad $(\hat{\vepsilon}_\mL, \hat{\vepsilon}_\mA, \hat{\vepsilon}_\mF) \leftarrow \phi(\mL_t, \mA_t, \mF_t, N, t)$
\STATE Calculate loss components:
\STATE \quad $\mathcal{L}_{\text{lattice}} \leftarrow \|\vepsilon_\mL - \hat{\vepsilon}_\mL\|^2_2$ 
\STATE \quad $\mathcal{L}_{\text{atom}} \leftarrow \|\vepsilon_\mA - \hat{\vepsilon}_\mA\|^2_2$
\STATE \quad $\mathcal{L}_{\text{coord}} \leftarrow \lambda_t\|\nabla_{\mF_t}\log q(\mF_t|\mF_0) - \hat{\vepsilon}_\mF\|^2_2$
\STATE Optimize network parameters by minimizing $\mathcal{L}_{\text{total}} = \mathcal{L}_{\text{lattice}} + \mathcal{L}_{\text{atom}} + \mathcal{L}_{\text{coord}}$
\end{algorithmic}
\end{algorithm}

\begin{algorithm}[H]
\small
\caption{Sampling Procedure for InvDesFlow-AL Pretrained Crystal Generation Model}\label{alg:gen_v2}
\begin{algorithmic}[1]
\REQUIRE Atom count $N$, denoising model $\phi$, diffusion steps $T$, Langevin step size $\gamma$
\ENSURE Generated crystal structure $(\mL_0, \mA_0, \mF_0)$
\STATE Initialize noise components:
\STATE \quad $\mL_T \sim \mathcal{N}(\mathbf{0},\mathbf{I})$, $\mA_T \sim \mathcal{N}(\mathbf{0},\mathbf{I})$, $\mF_T \sim \mathcal{U}(0,1)$
\STATE Compute noise schedule parameters $\{\alpha_t, \beta_t, \sigma_t\}_{t=1}^T$

\FOR{$t \gets T,\cdots, 1$}
    \STATE Sample process noises:
    \STATE \quad $\vepsilon_\mL,\vepsilon_\mA,\vepsilon_\mF \sim \mathcal{N}(\mathbf{0},\mathbf{I})$
    
    \STATE Predict denoising directions:
    \STATE \quad $(\hat{\vepsilon}_\mL, \hat{\vepsilon}_\mA, \hat{\vepsilon}_\mF) \leftarrow \phi(\mL_t, \mA_t, \mF_t, N, t)$
    
    \STATE Update lattice parameters:
    \STATE \quad $\mL_{t-1} \leftarrow \frac{1}{\sqrt{\alpha_t}}\left(\mL_t - \frac{\beta_t}{\sqrt{1-\bar{\alpha}_t}}\hat{\vepsilon}_\mL\right) + \sqrt{\beta_t\cdot\frac{1-\bar{\alpha}_{t-1}}{1-\bar{\alpha}_t}}\vepsilon_\mL$
    
    \STATE Update atom types:
    \STATE \quad $\mA_{t-1} \leftarrow \frac{1}{\sqrt{\alpha_t}}\left(\mA_t - \frac{\beta_t}{\sqrt{1-\bar{\alpha}_t}}\hat{\vepsilon}_\mA\right) + \sqrt{\beta_t\cdot\frac{1-\bar{\alpha}_{t-1}}{1-\bar{\alpha}_t}}\vepsilon_\mA$
    
    \STATE Intermediate coordinate prediction:
    \STATE \quad $\mF_{t-\frac{1}{2}} \leftarrow w\left(\mF_t + (\sigma_t^2-\sigma_{t-1}^2)\hat{\vepsilon}_\mF + \frac{\sigma_{t-1}\sqrt{\sigma_t^2-\sigma_{t-1}^2}}{\sigma_t}\vepsilon_\mF\right)$
    
    \STATE Refine coordinate prediction:
    \STATE \quad $(_, \hat{\vepsilon}_\mF') \leftarrow \phi(\mL_{t-1}, \mF_{t-\frac{1}{2}}, \mA_{t-1}, N, t-1)$
    \STATE \quad $d_t \leftarrow \gamma\sigma_{t-1}^2/\sigma_1^2$ \COMMENT{Adaptive step scaling}
    \STATE \quad $\mF_{t-1} \leftarrow w\left(\mF_{t-\frac{1}{2}} + d_t\hat{\vepsilon}_\mF' + \sqrt{2d_t}\vepsilon_\mF'\right)$
\ENDFOR

\STATE \textbf{Return} $(\mL_0, \mA_0, \mF_0)$
\end{algorithmic}
\end{algorithm}

\subsection*{Active learning-based materials inverse design}
In the generation of low formation energy materials, to select the most valuable data for iterative fine-tuning of the generative model in active learning, we designed a multi-objective scoring function to evaluate the newly generated crystal structures:
\begin{equation}
    \label{s-form}
    \mathcal{S}^\text{Form}_{\text{QBC}} = \underbrace{ (- E_{\text{form}})}_{\text{Stability}} 
    \; \cdot \;
    \underbrace{ (\mathbb{I}_{\text{relaxation}})}_{\text{Force convergence}} 
    \; \cdot \;
    \underbrace{  (\mathbb{I}_{\text{novelty}})}_{\text{New crystal}}.
\end{equation}
Similarly, in the generation of low E$_\text{hull}$ materials, the multi-objective scoring function for selecting the most valuable data is:
\begin{equation}
    \mathcal{S}^\text{Hull}_{\text{QBC}} = \underbrace{ (- E_{\text{hull}})}_{\text{Synthesizability}}  
    \; \cdot \;
    \underbrace{ (\mathbb{I}_{\text{relaxation}})}_{\text{Force convergence}}  
    \; \cdot \;
    \underbrace{ (\mathbb{I}_{\text{novelty}})}_{\text{New crystal}}.
\end{equation}
For the generation process of high-temperature superconducting materials, we devised a multi-objective scoring function to assess the newly generated crystal structures to identify the most valuable data for iterative fine-tuning of the generative model during active learning:
\begin{equation}
    \mathcal{S}^\text{Supercon}_{\text{QBC}} = \underbrace{T_c^{\text{DFT}}\quad \text{or} \quad T_c^{\text{SuperconGNN}}}_{\text{High-Tc priority}}  
    \; \cdot \;
    \underbrace{ (\mathbb{I}_{\text{relaxation}})}_{\text{Force convergence}}  
    \; \cdot \;
    \underbrace{ (\mathbb{I}_{\text{novelty}})}_{\text{New crystal}}
    \; \cdot \;
    \underbrace{ (\mathbb{I}_{E_{{\text{hull}}}<50\text{meV}})}_{\text{Synthesizability}},
\end{equation}
the high-Tc priority term directs the model to generate structures within the high-temperature superconducting material space, prioritizing the inclusion of DFT-confirmed superconducting phases. Additionally, it utilizes SuperconGNN predictions of high-\(T_c\) superconducting materials as a reference for selection. The force convergence term ensures structural stability, while the synthesizability term balances synthesizability. These conditions together form the criteria for selecting the most valuable data.
In the stable structure prediction task, the DPA-2 potential function optimizes the generated structures, and FormEGNN is used to select the crystals with the lowest formation energy. The multi-objective scoring function is given in Equation~\ref{s-form}.

\begin{figure*}[h!]
		\centering  
		\includegraphics[width=1.0\linewidth]{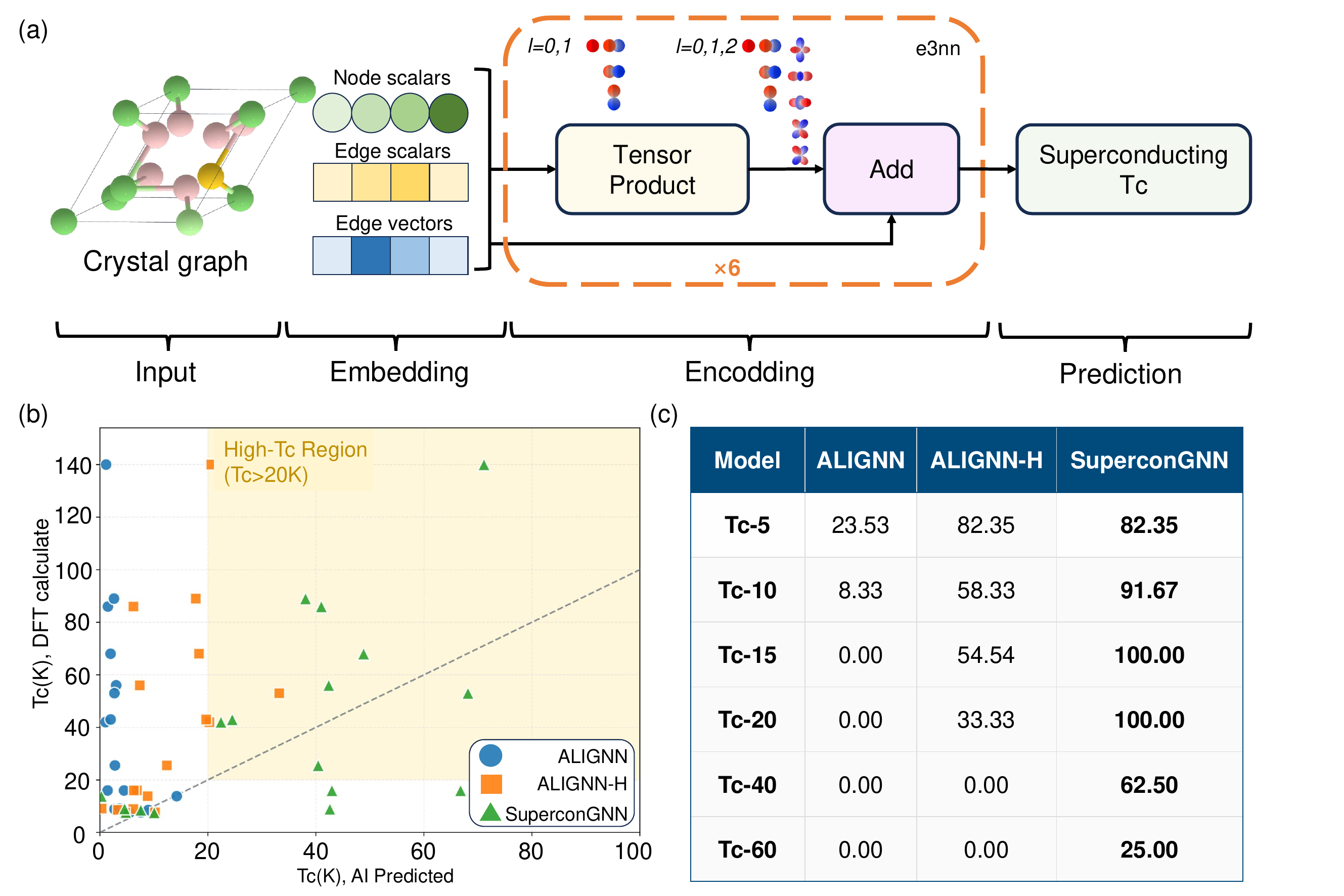}
		\caption{
        \textbf{SuperconGNN model architecture and performance.}  
(a) The input layer, Embedding layer, encoding layer, and prediction layer of the crystal graph in SuperconGNN.  
(b) Performance comparison of different models in predicting high-temperature superconducting materials.  
(c) Accuracy of AI predictions: whether AI can correctly classify materials with superconducting transition temperatures above 5K, 10K, ..., up to 60K into their corresponding intervals.
        }
		\label{SuperconGNN} 
\end{figure*}
SuperconGNN is a superconducting transition temperature prediction model we developed based on an equivariant graph neural network architecture. As shown in Figure~\ref{SuperconGNN} (a), the model consists of a crystal graph input layer, a embedding layer, an encoding layer, and a prediction layer.

The crystal structure is represented as a geometric graph $(\mathcal{V}, \mathcal{E}_{r_{\max} })$, where nodes correspond to atoms within the unit cell. We employ pymatgen~\footnote{https://pymatgen.org/} to construct a local environment graph by identifying bonding interactions between atoms. 
The crystallographic information file (CIF) of a structure is then converted into a graph representation that includes atomic types $\mA$, bonding connectivity $\mathcal{E}$, lattice parameters $\mL$, and periodic boundary information. In addition to the bonding-based edges obtained from the local environment analysis, we further construct a geometric graph based on atomic cartesian coordinates by connecting atoms within a cutoff distance of 5Å. These distance-based connections are subsequently merged with the bonding-based edges to form the final geometric graph representation:
\begin{equation}
\begin{gathered}
    \mathcal{E}_{r_{\max} } = \mathcal{E} \sqcup \{ (a,b) \mid r_{ab} < r_{\max} \}  \\
e_{ab} = \text{MLP}^{(e)}(f_{ab} || \mu(r_{ab})  \quad \forall (a,b) \in \mathcal{E}_{r_{\max}}  \\
V_a^{(0)} = \text{MLP}^{(v)}(f_a)  \quad \forall a \in \mathcal{V} 
\end{gathered}
\end{equation}
$r_{ab}$ denotes the Euclidean distance between atoms $a$ and $b$, $f_a$ represents the node scalar features, and $f_{ab}$ corresponds to the edge scalar features of the bond $(a, b)$ if it belongs to $\mathcal{E}$, and is 0 otherwise.
The node features $f_a$ consist of a one-hot encoding of the atomic type combined with global lattice information $(\alpha, \beta, \gamma, a, b, c)$. The edge features include a 2D one-hot encoding indicating the presence or absence of a bond, along with the corresponding gaussian-expanded interatomic distance.

The Encoding layers are based on tensor product layers~\cite{e3nn,diffdock,torsionaldiff}. At each layer, messages are generated for every node pair in the graph by applying tensor products between the current node features and the spherical harmonic representation of the corresponding normalized edge vector. The weights for these tensor products are computed as a function of the edge embeddings and the scalar features of the two connected nodes, where the scalar features of node \(a\) are denoted as \(\mathbf{h}^0_a\). These messages are then aggregated at each node, and the resulting information is used to update its feature representation.

\begin{equation}
\begin{gathered}
\mathbf{h}_a \leftarrow \mathbf{h}_a {\oplus} \textsc{BN}^{(a)} \Bigg( \frac{1}{|\mathcal{N}_a|}\sum_{b \in \mathcal{N}_a} Y(\hat r_{ab}) \; \otimes_{\psi_{ab}} \; \mathbf{h}_b \Bigg) \\
\text{with} \; \psi_{ab} = \Psi^{(a)}(e_{ab}, \mathbf{h}^0_a, \mathbf{h}^0_b)
\end{gathered}
\end{equation}
Here, \(\mathcal{N}_a = \{b \mid (a,b) \in \mathcal{E}_{r_{\text{max}}}\}\) denotes the set of neighboring atoms of atom \(a\), where \(\mathcal{E}_{r_{\text{max}}}\) refers to the set of edges within a predefined cutoff radius. \(Y\) represents the spherical harmonics up to order \(\ell = 2\), and \(\textsc{BN}\) indicates an equivariant batch normalization layer. All learnable parameters are encapsulated in \(\Psi\), which governs the weighting of tensor products. The resulting node features \(\mathbf{h}_a\) include both scalar and vector representations.
Considering that the superconducting transition temperature of a crystal is invariant under SE(3) transformations, the final layer of our model employs an SE(3)-invariant linear layer from e3nn to map the node features of the crystal to an SE(3)-invariant representation. Additionally, since \(T_c\) is strictly greater than zero, we apply a ReLU activation function. Finally, the invariant representations of all atoms are aggregated to predict the superconducting transition temperature.

The discovery of high-temperature superconducting materials has long been a central goal in condensed matter physics. Accordingly, we focus on evaluating the model's performance in the high-\(T_c\) regime. The yellow region in Figure~\ref{SuperconGNN}(b) illustrates the scatter plot of predicted values versus DFT-calculated values for materials with \(T_c > 20\,\text{K}\), as predicted by ALIGNN~\cite{Choudhary2022}, ALIGNN-H~\cite{MM-Searching}, and our model SuperconGNN. A greater number of points deviating from the diagonal reference line indicates superior model performance. As shown, SuperconGNN successfully predicts all superconducting materials with \(T_c > 20\,\text{K}\) within the highlighted region. The baseline model ALIGNN was trained on conventional superconductor datasets, whereas ALIGNN-H was trained specifically on hydride data.
Due to variations in DFT computational settings, theoretical predictions of superconducting transition temperature (\(T_c\)) can exhibit discrepancies exceeding 10 K. Moreover, differing considerations of factors such as anisotropy among experts can lead to even larger deviations for the same material, with reported \(T_c\) values differing by more than 50 K—for example, Mg\(_2\)IrH\(_6\) has been predicted to exhibit a \(T_c\) of both 160 K~\cite{mg2irh6-160k} and 77 K~\cite{mg2irh6-77k}. Therefore, an excessive focus on reducing the AI prediction error (e.g., MAE below 10 K) is not meaningful in the context of discovering high-\(T_c\) superconductors.  
To address this, we propose a novel evaluation criterion, as illustrated in Figure~\ref{SuperconGNN}(c). Specifically, a prediction is considered correct if the model successfully predicts \(T_c\) above a given threshold, which allows us to compute an accuracy score that better reflects the model’s capability in the high-\(T_c\) regime. Using this approach, we observe that SuperconGNN accurately predicts materials with \(T_c > 15\) K and \(T_c > 20\) K, significantly outperforming the baseline models. In the regime beyond the McMillan limit, SuperconGNN also achieves high predictive accuracy. Overall, SuperconGNN offers a novel, efficient, and accurate model for the discovery of high-\(T_c\) superconductors.

Regarding data partitioning, we follow the same strategy described in Ref.~\cite{Choudhary2022}, which involves splitting the 626 conventional superconductor data points into training, validation, and test sets with a ratio of 0.9:0.05:0.05. Since the specific data identifiers were not released by the authors, we performed a new split using the same proportions. Given that the majority of these data points correspond to materials with superconducting transition temperatures below 40 K, we further augmented the training set with 59 hydride crystal structures reported in Ref.~\cite{MM-Searching}. We directly adopted the final checkpoint of our model as the screening model. To evaluate its screening capability, we added 12 superconducting candidates discovered by InvDesFlow-AL to the test set and used them to assess model performance in the corresponding figures.

We discuss the differences between InvDesFlow-AL and current reinforcement learning approaches. While reinforcement learning has been extensively applied in domains such as large language models and robotic control, achieving remarkable outcomes like GPT4~\cite{openai2024gpt4technicalreport} and DeepSeek~\cite{deepseekai}, methods like reinforcement learning from human feedback (RLHF)~\cite{RLHF-GPT} require proximal policy optimization (PPO)~\cite{RL-PPO} and additional reward model training, resulting in high computational complexity and training instability. Our active learning strategy resembles the direct preference optimization (DPO)~\cite{RL-DPO} in reinforcement learning by fine-tuning models directly with preference data, which offers lower computational overhead and eliminates the need for separate reward modeling. The distinction between InvDesFlow-AL and DPO lies in the former's elimination of negative sampling, instead directly updating the model towards the distribution of preferred data generation.

\subsection*{The first-principles electronic calculation.}
As shown in Table~\ref{tab:calculation_parameters}, this includes the DFT Settings, EPC Calculation, and the corresponding BCS Theory for Li$_2$AuH$_6$. As shown in Figure~\ref{invdesflow-AL-supercon} (a), InvDesFlow-AL also identified many other superconducting materials, with the corresponding DFT calculation results provided in the SM.Sec-A.
\begin{table}[htbp]
\caption{Summary of DFT computational parameters and electron-phonon coupling methodology.}
\label{tab:calculation_parameters}
\centering
\begin{tabular}{lll}
\toprule
\textbf{Category} & \textbf{Parameter/Method} & \textbf{Specification} \\
\midrule
\multirow{6}{*}{DFT Settings} 
& Software package & QUANTUM-ESPRESSO~\cite{Giannozzi_2009} \\
& Exchange-correlation & PBE-GGA~\cite{PhysRevLett.77.3865} \\
& Pseudopotentials & Optimized norm-conserving Vanderbilt~\cite{PhysRevB.88.085117} \\
& Cutoff energies & Kinetic: 80 Ry, Charge density: 320 Ry \\
& $\mathbf{k}$-mesh & 16$\times$16$\times$16 (unshifted) \\
& Smearing & Methfessel-Paxton (0.02 Ry)~\cite{PhysRevB.40.3616} \\
& Phonon calculation & 4$\times$4$\times$4 $\mathbf{q}$-mesh (DFPT)~\cite{RevModPhys.73.515} \\
\midrule
\multirow{5}{*}{EPC Calculation}
& Methodology & Wannier interpolation (EPW)~\cite{PONCE2016116} \\
& Wannier functions & MLWFs~\cite{Pizzi_2020} on 4$\times$4$\times$4 $\mathbf{k}$-mesh \\
& Projected orbitals & Au-5d, H-1s \\
& Fine grids & Electron: 48$\times$48$\times$48, Phonon: 16$\times$16$\times$16 \\
& Smearing widths & 90 meV (electron), 0.5 meV (phonon) \\
& Eliashberg solver & Anisotropic equations~\cite{PONCE2016116} \\
& Matsubara cutoff & $\omega_c = 1.7$ eV (10$\times$ max phonon freq) \\
\midrule
\multirow{3}{*}{Theory}
& EPC constant & $\lambda = \frac{1}{N_\mathbf{q}} \sum_{\mathbf{q}\nu} \lambda_{\mathbf{q}\nu} = 2 \int \frac{\alpha^2 F(\omega)}{\omega} d\omega$ \\
& Spectral function & $\alpha^2 F(\omega) = \frac{1}{2N_\mathbf{q}} \sum_{\mathbf{q}\nu} \lambda_{\mathbf{q}\nu} \omega_{\mathbf{q}\nu} \delta(\omega-\omega_{\mathbf{q}\nu})$ \\
& Coupling formula & $\lambda_{\mathbf{q}\nu} = \frac{2}{\hbar N(0) N_\mathbf{k}} \sum_{nm \mathbf{k}} \frac{|g^{nm}_{\mathbf{k,q}\nu}|^2}{\omega_{\mathbf{q}\nu}} \delta(\epsilon^n_{\mathbf{k}}) \delta(\epsilon^m_{\mathbf{k+q}})$ \\
\bottomrule
\end{tabular}
\end{table}

\section*{Data availability} 
The crystal data are available from the Materials Project database via the web interface at \url{https://materialsproject.org} or the API at \url{https://api.materialsproject.org}.

\section*{Code availability} 
All the data, code~(\url{https://github.com/xqh19970407/InvDesFlow-AL}), and models will be open-sourced after the paper is published.
We rely on PyTorch~(\url{https://pytorch.org}) for deep model training.
We use specialized tools for the Vienna Abinitio Simulation Package~(\url{https://www.vasp.at/}).

\vspace{36pt}
\noindent\textbf{Acknowledgement:}
The work is supported by the National Natural Science Foundation of China (No.62476278, No.11934020). Computational resources have been provided by the Physical Laboratory of High Performance Computing at Renmin University of China.
\\

\noindent\textbf{Corresponding authors:} Correspondence and requests for materials should be addressed to Ze-Feng Gao (zfgao@ruc.edu.cn), Hao Sun (haosun@ruc.edu.cn) and Zhong-Yi Lu (zlu@ruc.edu.cn). \\

\noindent\textbf{Competing interests:}
The authors declare no competing interests.\\

\noindent\textbf{Supplementary information:}
The supplementary information is attached.

\bibliographystyle{unsrt}
\bibliography{references}

\clearpage

\noindent This supplementary document provides a detailed description of the proposed pre-trained model, dataset statistics, hyperparameter value, and details of altermagnetic materials confirmed by electronic structure calculations.

\section{DFT calculation results}
Figures S.1–S.3 systematically present the structural and physical properties of a series of InvDesFlow-AL discovered superconductors. In each figure, panels (a) and (b) show the crystal structures of two representative materials: Ca$_2$CuH$_6$ and K$_2$GaCuH$_6$ in Figure S.1, K$_2$CdCuH$_6$ and K$_2$LiZnH$_6$ in Figure S.2, and Na$_2$GaCuH$_6$ and Na$_2$LiAgH$_6$ in Figure S.3. Panels (c) and (d) of each figure present the corresponding phonon spectra, phonon density of states (DOS), electronic band structures, and electronic density of states, thereby providing a comprehensive view of both the lattice dynamics and electronic properties of these compounds.

\begin{figure*}[h]
		\centering  
		\includegraphics[width=1.0\linewidth]{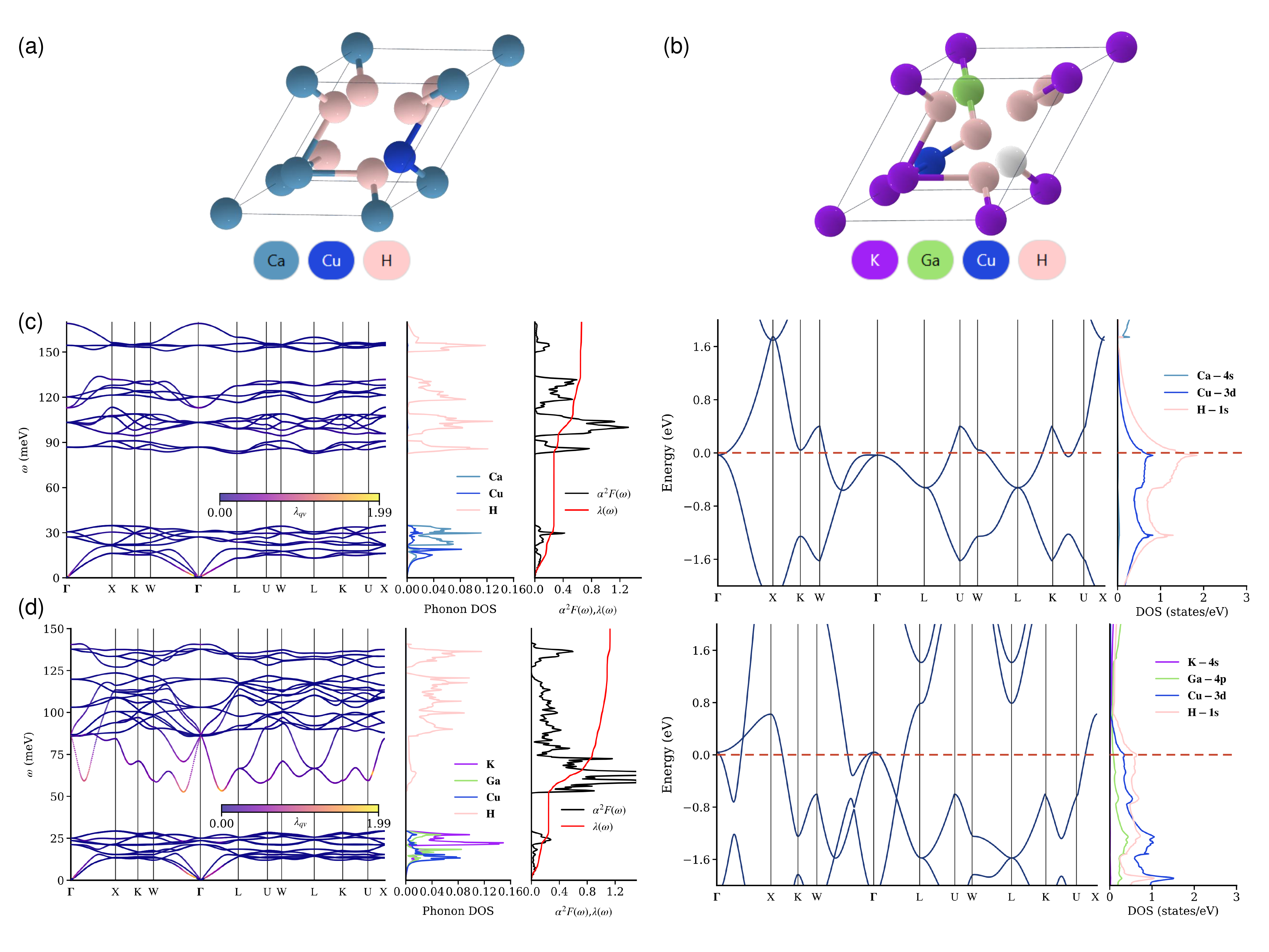}
		\caption{
            (a) and (b) present the crystal structures of Ca$_2$CuH$_6$ and K$_2$GaCuH$_6$, respectively.  
(c) displays the phonon spectrum, phonon density of states, electronic band structure, and density of states for Ca$_2$CuH$_6$.  
(d) shows the corresponding phonon spectrum, phonon density of states, electronic band structure, and density of states for K$_2$GaCuH$_6$.  
        }
		\label{Ca2CuH6-band-2} 
\end{figure*}

\begin{figure*}[h]
		\centering  
		\includegraphics[width=1.0\linewidth]{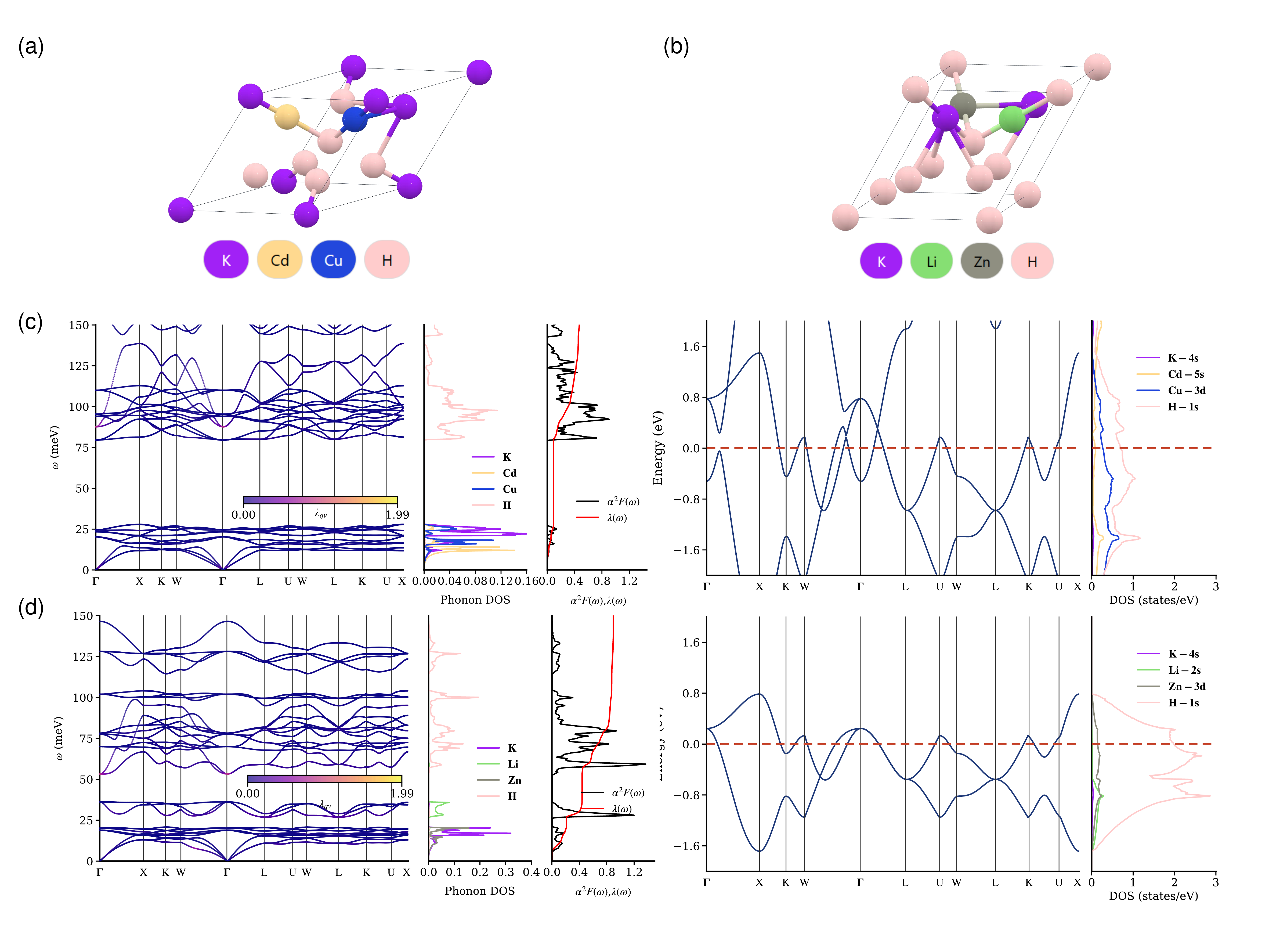}
		\caption{
            (a) and (b) present the crystal structures of K$_2$CdCuH$_6$ and K$_2$LiZnH$_6$, respectively.  
(c) displays the phonon spectrum, phonon density of states, electronic band structure, and density of states for K$_2$CdCuH$_6$.  
(d) shows the corresponding phonon spectrum, phonon density of states, electronic band structure, and density of states for K$_2$LiZnH$_6$.  
        }
		\label{Ca2CuH6-band-3} 
\end{figure*}

\begin{figure*}[h]
		\centering  
		\includegraphics[width=1.0\linewidth]{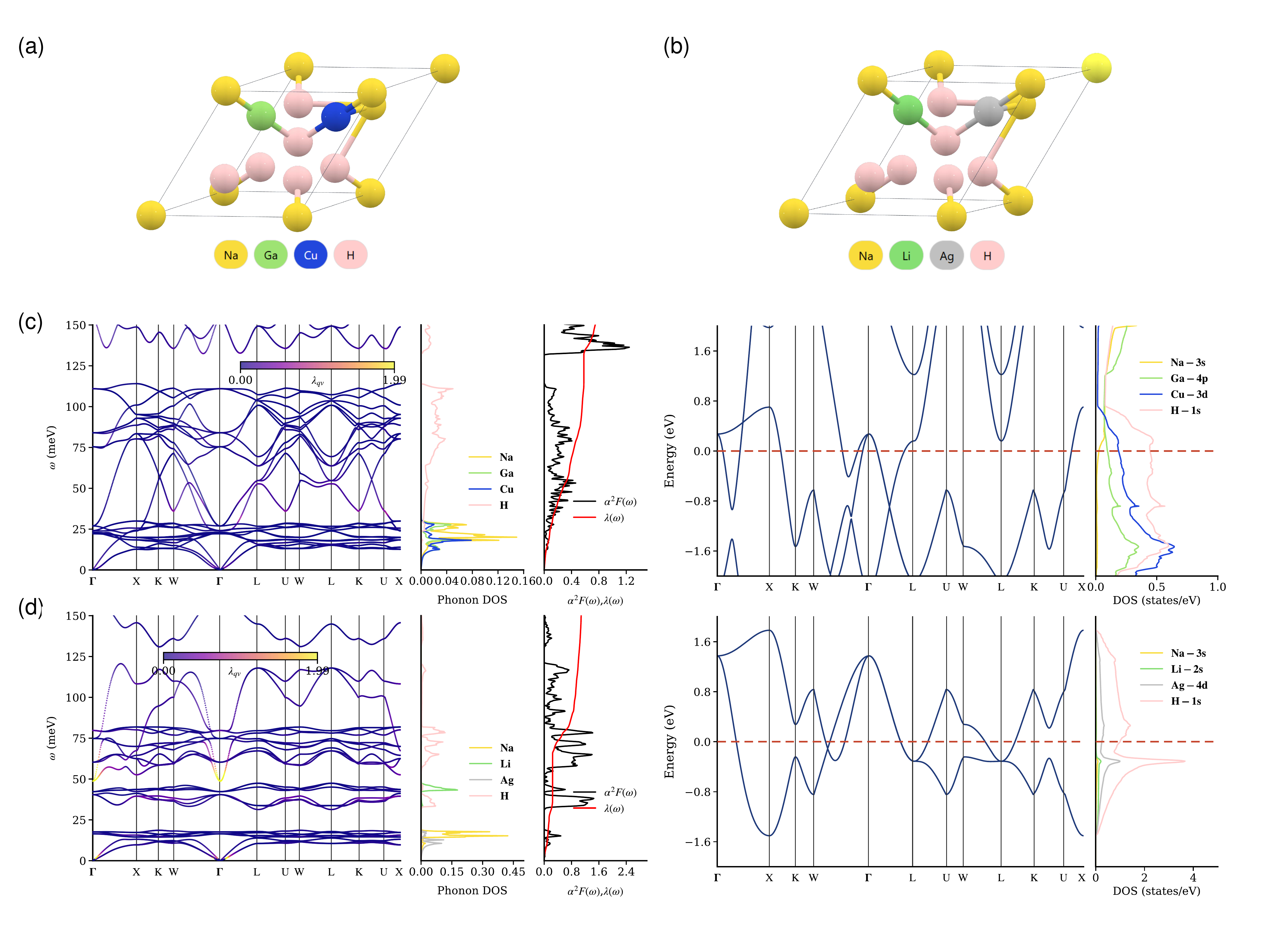}
		\caption{
            (a) and (b) present the crystal structures of Na$_2$GaCuH$_6$ and Na$_2$LiAgH$_6$, respectively.  
(c) displays the phonon spectrum, phonon density of states, electronic band structure, and density of states for Na$_2$GaCuH$_6$.  
(d) shows the corresponding phonon spectrum, phonon density of states, electronic band structure, and density of states for Na$_2$LiAgH$_6$. 
        }
		\label{Ca2CuH6-band} 
\end{figure*}

\begin{table}[ht]
\centering
\caption{Comparison between predicted $T_c$ values by SuperconGNN and those calculated by solving the anisotropic Eliashberg equations.}
\label{table:Tc}
\begin{tabular}{lcc}
\toprule
\textbf{Formula} & \textbf{SuperconGNN} & \textbf{$T_c$ (K)} \\
\midrule
Li$_2$AuH$_6$ & 71 & 140 \\
Na$_2$LiAgH$_6$ & 41 & 86 \\
K$_2$GaCuH$_6$ & 48 & 68 \\
Na$_2$GaCuH$_6$ & 22 & 42 \\
K$_2$LiZnH$_6$ & 40 & 25 \\
Ca$_2$CuH$_6$ & 42 & 16 \\
K$_2$CdCuH$_6$ & 42 & 9 \\
\bottomrule
\end{tabular}
\end{table}

\section{Training Details and Hyperparameter Settings}
Table \ref{tab:pre-train_params} outlines the training configuration for the pre-trained crystal generation model. The data pipeline utilizes CrystalNN-based graph construction with lattice scaling, and each structure contains up to 20 atoms. The model adopts an EGNN-based decoder with 6 graph neural network layers, a hidden dimension of 512, and SiLU activation. The training objective is governed by a diffusion process with 1,000 steps and weighted losses for coordinates, lattice parameters, and atom types. A neighborhood cutoff of 7.0 Å and a maximum of 20 neighbors per atom are used to define atomic interactions. Optimization is performed using the Adam optimizer with a base learning rate of \(1 \times 10^{-4}\), and learning rate scheduling is managed via ReduceLROnPlateau. The model is trained for 1,000 epochs on an NVIDIA GeForce RTX 4090 GPU. Full hyperparameter settings are provided in Table \ref{tab:pre-train_params}.

\begin{table}[ht]
\centering
\caption{Training configuration details of pre-trained crystal generation mode.}
\label{tab:pre-train_params}
\begin{tabular}{ll}
\toprule
\textbf{Category} & \textbf{Parameter Settings} \\
\midrule
\textbf{Data Configuration} & \\
\quad Graph construction & CrystalNN method with lattice scaling \\
\quad Max atoms per structure & 20 \\
\quad Tolerance for structure matching & 0.1 \\
\quad Train/Val/Test batch sizes & 96/64/64 \\
\quad Parallel preprocessing workers & 30 \\

\textbf{Model Architecture} & \\
\quad Decoder type & EGNN \\
\quad Hidden dimension & 512 \\
\quad Number of GNN layers & 6 \\
\quad Activation function & SiLU \\
\quad Diffusion steps & 1000 \\
\quad Cost weights (coord/lattice/type) & 1.0/1.0/20.0 \\
\quad Neighborhood cutoff radius & 7.0 Å \\
\quad Max neighbors per atom & 20 \\

\textbf{Optimization} & \\
\quad Optimizer & Adam \\
\quad Base learning rate & $1 \times 10^{-4}$ \\
\quad Learning rate scheduler & ReduceLROnPlateau (factor=0.6, patience=30) \\
\quad Minimum learning rate & $1 \times 10^{-4}$ \\

\textbf{Training Protocol} & \\
\quad Epochs & 1000 \\

\textbf{Hardware \& Logging} & \\
\quad GPU device & NVIDIA GeForce RTX 4090 \\
\bottomrule
\end{tabular}
\end{table}

Table \ref{tab:supercongnn_params} presents the training configuration for the SuperconGNN model, which is based on spherical harmonics and designed for predicting superconducting transition temperatures. The data processing utilizes a radius graph construction with a maximum radius of 5.0, and input edge features are set to 2. The model consists of 128 scalar features, 10 vector features, and employs a spherical harmonics expansion up to the second order. It uses 6 convolutional layers with ReLU activation and incorporates third-order representations and residual connections. The optimization is carried out with the Adam optimizer, a base learning rate of \(1 \times 10^{-4}\), and a warmup linear decay learning rate scheduler, where the warmup steps are 0.5 times the total number of steps. The model is trained for 200 epochs with a mean squared error loss function on an NVIDIA GeForce RTX 4090 GPU. Detailed hyperparameter settings are provided in Table \ref{tab:supercongnn_params}.

\begin{table}[ht]
\centering
\caption{Training configuration details of SuperconGNN.}
\label{tab:supercongnn_params}
\begin{tabular}{ll}
\toprule
\textbf{Category} & \textbf{Parameter Settings} \\
\midrule
\textbf{Data Configuration} & \\

\quad Graph construction & Radius Graph \\
\quad Maximum radius & 5.0 \\
\quad Input edge features & 2 \\
\quad Train/Val/Test batch sizes & 32/32/32 \\

\textbf{Model Architecture} & \\
\quad Model type & SuperconGNN \\
\quad Number of scalar features (ns) & 128 \\
\quad Number of vector features (nv) & 10 \\
\quad Maximum spherical harmonics order (sh\_lmax) & 2 \\
\quad Number of convolutional layers & 6 \\
\quad Activation function & ReLU \\
\quad Use third-order representation & True \\
\quad Residual connections & True \\

\textbf{Optimization} & \\
\quad Optimizer & Adam \\
\quad Base learning rate & $1 \times 10^{-4}$ \\
\quad Learning rate scheduler & Warmup Linear Decay \\
\quad Warmup steps & 0.5 $\times \text{Total Steps}$ \\

\quad Minimum learning rate & $1 \times 10^{-4}$ \\

\textbf{Training Protocol} & \\
\quad Number of epochs & 200 \\
\quad Loss function & Mean Squared Error\\

\textbf{Hardware \& Logging} & \\
\quad GPU device & NVIDIA GeForce RTX 4090 \\
\bottomrule
\end{tabular}
\end{table}

\section{Other Visualizations}
\begin{figure*}[h]
		\centering  
		\includegraphics[width=0.7\linewidth]{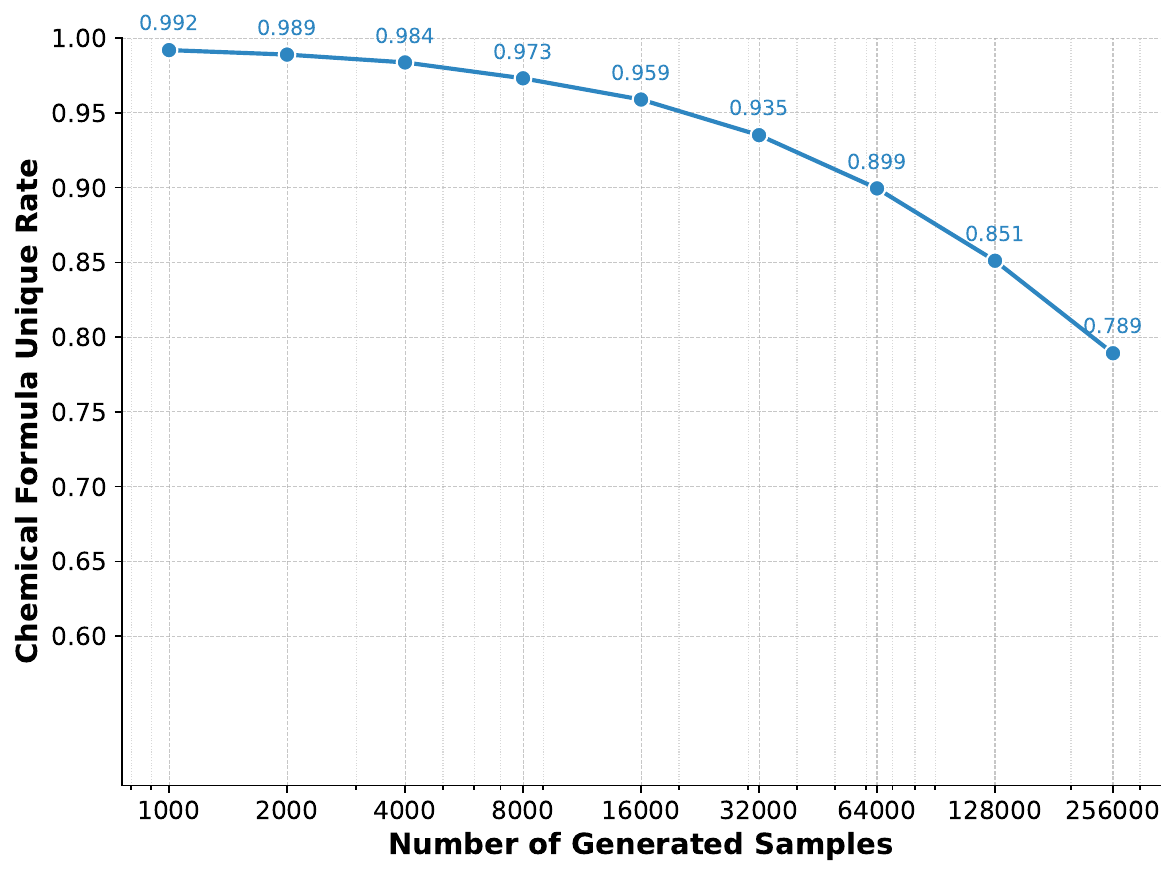}
		\caption{The ratio of unique chemical formulas when the InvDesFlow-AL pre-trained generative model generates 1,000, 2,000, 4,000, 8,000, 16,000, 32,000, 64,000, 128,000, and 256,000 materials.
        }
		\label{unique-rate-trend} 
\end{figure*}

\begin{figure*}[h]
		\centering  
		\includegraphics[width=0.7\linewidth]{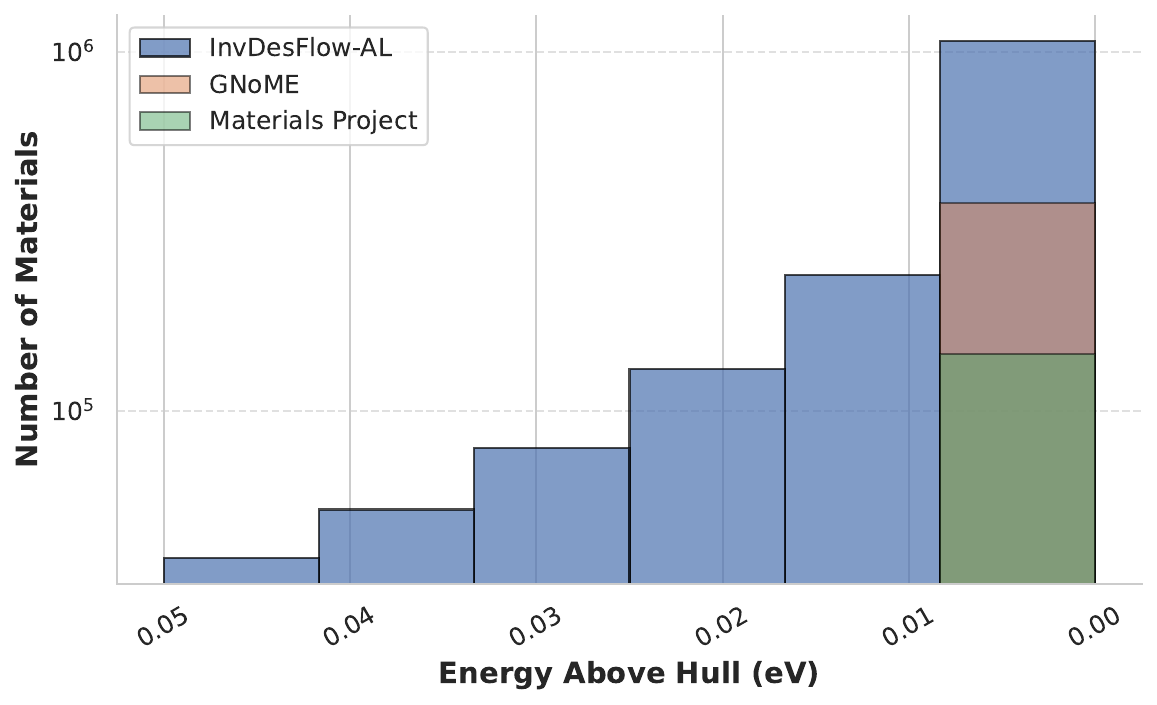}
		\caption{
            A histogram of the 1,598,551 materials with Ehull < 50 meV generated by InvDesFlow-AL, categorized by different Ehull intervals.
        }
		\label{invdesflow-AL-hull50} 
\end{figure*}

As shown in Figure~\ref{unique-rate-trend}, the ratio of unique chemical formulas generated by the InvDesFlow-AL pre-trained generative model is reported for 1,000, 2,000, 4,000, 8,000, 16,000, 32,000, 64,000, 128,000, and 256,000 materials. As the number of generated samples increases, the uniqueness ratio gradually declines. Nevertheless, even when generating over 256,000 materials, the model still demonstrates a high level of novelty and uniqueness.
Figure~\ref{invdesflow-AL-hull50} presents a statistical overview and specific case analysis of the materials generated by InvDesFlow-AL. In panel (a), a histogram illustrates the distribution of 1,598,551 generated materials with formation enthalpy (Ehull) less than 50 meV, categorized by different Ehull intervals. This distribution highlights the thermodynamic stability landscape of the generated compounds, with a significant number of candidates falling within low-Ehull regions, suggesting good potential for synthesizability.
Figure~\ref{invdesflow-AL-supercon} (continued) showcases a case study of Zn$_3$Au$_4$, one of the promising candidates generated. Panel (a) displays its binary phase diagram, offering insight into its thermodynamic compatibility and phase coexistence with related compounds. Panel (b) illustrates the crystal structure of Zn$_3$Au$_4$, revealing its atomic arrangement and symmetry.

\begin{figure*}[h]
		\centering  
		\includegraphics[width=0.9\linewidth]{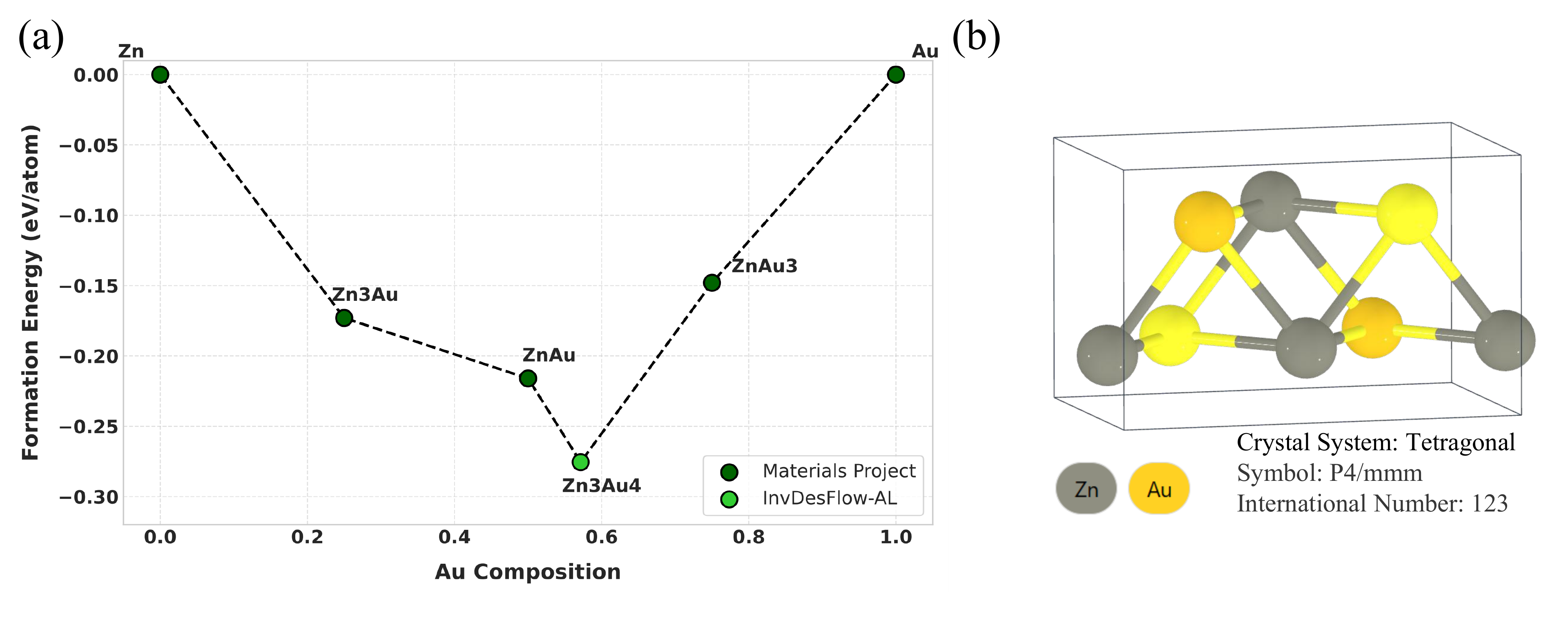}
		\caption{
            (a) Binary phase diagram of Zn$_3$Au$_4$  
(b) Crystal structure of Zn$_3$Au$_4$
        }
		\label{invdesflow-AL-supercon} 
\end{figure*}

\footnotesize

\end{document}